\documentclass{article}

\usepackage{amsmath}
\usepackage{amssymb}
\usepackage{acronym}
\usepackage{algorithm,algpseudocode}
\usepackage{amsfonts}       % blackboard math symbols
\usepackage{booktabs}       % professional-quality tables
\usepackage{caption}
\usepackage{comment}
\usepackage[T1]{fontenc}    % use 8-bit T1 fonts
\usepackage{fancyhdr}       % header
\usepackage{graphicx}       % graphics
\usepackage{hyperref}
\usepackage{lipsum}
\usepackage{multirow}
\usepackage{mathrsfs}
\usepackage{microtype}      % microtypography
\usepackage{nicefrac}       % compact symbols for 1/2, etc.
\usepackage[utf8]{inputenc} % allow utf-8 input
\usepackage{PRIMEarxiv}
\usepackage{subcaption}
\usepackage{url}            % simple URL typesetting
\usepackage[table]{xcolor}

\graphicspath{{media/}}     % organize your images and other figures under media/ folder
\hypersetup{
    colorlinks=true,
    allcolors=blue,
    }

\newtheorem{theorem}{Theorem}

\def\R{\mathbb{R}}
\def\E{\mathbb{E}}

\def\P{\mathrm{P}}

\def\eps{\epsilon}

\providecommand{\nnorm}[1]{ \lVert#1 \rVert}
\providecommand{\norm}[1]{ \left \lVert#1 \right \rVert}
\def \coloneq{\mathrel{\mathop:}=}
\providecommand{\M}[1]{\mathbf#1}
\usepackage{bm}
\providecommand{\T}{\top}
\DeclareMathAlphabet\mathbfcal{OMS}{cmsy}{b}{n}
\providecommand{\wh}[1]{\widehat{#1}}

\newcommand{\nscp}[2]{\langle#1, #2\rangle}
\usepackage[normalem]{ulem} 
{\makeatletter
 \gdef\xxxmark{%
   \expandafter\ifx\csname @mpargs\endcsname\relax % in minipage?
     \expandafter\ifx\csname @captype\endcsname\relax % in figure/caption?
       \marginpar{\textcolor{red}{xxx~}}% not in a caption or minipage, can use marginpar
     \else
       \textcolor{red}{xxx~}% notice trailing space
     \fi
   \else
     \textcolor{red}{xxx~}% notice trailing space
   \fi}
 \gdef\xxx{\@ifnextchar[\xxx@lab\xxx@nolab}
 \long\gdef\xxx@lab[#1]#2{{\bf [\xxxmark \textcolor{red}{#2} ---{\sc #1}]}}
 \long\gdef\xxx@nolab#1{{\bf [\xxxmark \textcolor{red}{#1}]}}
}

\title{Tensor Completion for Causal Inference with Multivariate Longitudinal Data: A Reevaluation of COVID-19 Mandates} 

\author{
  Jonathan Auerbach, \, Martin Slawski, \, and Shixue Zhang \\
  Department of Statistics \\
  George Mason University \\
  Fairfax, Virginia\\
  \texttt{jauerba@gmu.edu} \\
}

\begin{document}
\maketitle

\begin{abstract} 
We propose a new method that uses tensor completion to estimate causal effects with multivariate longitudinal data---data in which multiple outcomes are observed for each unit and time period. Our motivation is to estimate the number of COVID-19 fatalities prevented by government mandates such as travel restrictions, mask-wearing directives, and vaccination requirements. In addition to COVID-19 fatalities, we observe related outcomes such as the number of fatalities from other diseases and injuries. The proposed method arranges the data as a tensor with three dimensions---unit, time, and outcome---and uses tensor completion to impute the missing counterfactual outcomes. We first prove that under general conditions, combining multiple outcomes using the proposed method improves the accuracy of counterfactual imputations. We then compare the proposed method to other approaches commonly used to evaluate COVID-19 mandates. Our main finding is that other approaches overestimate the effect of masking-wearing directives and that mask-wearing directives were not an effective alternative to travel restrictions. We conclude that while the proposed method can be applied whenever multivariate longitudinal data are available, we believe it is particularly timely as governments increasingly rely on longitudinal data to choose among policies such as mandates during public health emergencies. 
\end{abstract}

\section{Introduction}
\label{sec:heading}

This paper proposes a new approach to estimating causal effects with multivariate longitudinal data. Our motivation is to evaluate government mandates adopted during the pandemic, such as travel restrictions, mask-wearing directives, and vaccine requirements. Our evaluation follows Rubin, Holland, and Neyman \cite{holland1986statistics, rubin2005causal} in that we regard estimating the effect of a mandate as tantamount to predicting the number of COVID-19 fatalities that would have occurred had a government not adopted it. In other words, we observe the number of fatalities that occur in a jurisdiction while a mandate is in place, and this paper is concerned with predicting the number of fatalities that would have occurred had the mandate not been in place. The main difference between our approach and other COVID-19 analyses is that we use multiple outcomes to make these predictions---not just the number of COVID-19 fatalities from other jurisdictions and time periods. These additional outcomes are not of direct interest themselves, but they provide valuable information about COVID-19 fatalities.

Our main insight is that tensor completion provides a simple and effective framework for combining multiple outcomes across multiple jurisdictions. The proposed approach generalizes matrix completion, which has proven to be a versatile tool for making predictions with matrix-oriented data, such as univariate longitudinal data. An analysis with the proposed approach works like most longitudinal analyses: It leverages the staggered adoption of a mandate to borrow from the experience of control jurisdictions, whose governments did not contemporaneously adopt the mandate. However, unlike most longitudinal analyses, it also borrows from the experience of control outcomes that would not otherwise be of interest.

The approach is particularly useful when there are few time periods, when one jurisdiction adopts a mandate for many time periods, or when many jurisdictions adopt a mandate in the same time period. These cases are not uncommon---all three occur when evaluating COVID-19 mandates---and when they occur, traditional approaches can fail because a single outcome provides insufficient overlap between jurisdictions and time periods. In order to borrow correctly from the experience of control jurisdictions, univariate analyses rely on strong parametric assumptions that cannot be validated from the data.

By including additional control outcomes, we increase the overlap between jurisdictions and time periods, reducing the reliance on untestable parametric assumptions. The additional outcomes are not of direct interest themselves. Rather, they are informative of unobserved factors that influence the outcome of interest. For example, many U.S. states adopted travel restrictions early in the pandemic to reduce tourism, a factor contributing to the spread of COVID-19. A univariate longitudinal analysis would borrow from the experience of control states that did not reduce tourism through the use of travel restrictions. But tourism in the states that elected to restrict travel is not the same as tourism in states that did not, and the pandemic has not been observed over enough seasons to establish the difference. In contrast, a multivariate longitudinal analysis might borrow from the experience of other diseases spread by tourism, like influenza, which, unlike COVID-19, have been observed over many seasons, and thus can help account for the differential impact of tourism across states.

The idea of comparing multiple causes of death dates back to the seventeenth century \cite{willcox1938founder, graunt1939natural}, suggesting it is perhaps obvious that additional outcomes improve predictions when included judiciously. Yet multivariate longitudinal data have not become widely utilized, we suspect, because it is often not obvious how to include additional outcomes judiciously. Policymakers cannot fix outcomes like the conditions of an experiment. Rather, outcomes are measured together with error or influenced by similar unobserved factors, and thus they should be treated as jointly random. This is problematic when the joint distribution of the outcomes is unknown or intractable, and a reasonable parametric model cannot be specified easily. Such is the case when studying the government mandates adopted during the COVID-19 pandemic.

It is for this reason that we propose researchers arrange multivariate longitudinal data as a tensor and use tensor completion to estimate causal effects. We argue the approach does not make as strong parametric assumptions about the relationship between jurisdictions and time periods, can more accurately measure the effect of the government mandates adopted during the COVID-19 pandemic, and therefore should be considered as an additional tool for studying the pandemic. We present our argument in three sections. Section \ref{sec:back} introduces the proposed approach, provides a theoretical justification, and outlines inference. Section \ref{sec:covid} reevaluates three mandates adopted by U.S. states during the COVID-19 pandemic. Section \ref{sec:discussion} concludes with a discussion. The proof of the main theorem and a simulation study are provided in the Appendix. Our main conclusion is that while the proposed approach can be applied whenever multivariate longitudinal data are available, we believe it is particularly timely as governments increasingly rely on longitudinal data to choose among policies such as mandates during public health emergencies. 

\section{Tensor Completion for Causal Inference with Multivariate Longitudinal Data}
\label{sec:back}

Matrix completion problems have received considerable attention in machine learning, optimization, and high-dimensional statistics \cite{Rennie2005, Koren2009, Candes2009, Candes2010, Koltchinskii2011, Negahban2012, Mazumder2010, Ge2016}. While much of the attention comes from applications in collaborative filtering and recommendation systems---following the now famous solution to the "Netflix problem"---matrix completion has proven versatile---solving missing data problems in a wide variety of settings \cite{Gao2016, Li2020, Gu2014, Natarajan2014, Descary2019, Alidaee2020}. The potential outcome framework initiated by Rubin, Holland, and Neyman \cite{holland1986statistics, rubin2005causal} reduces causal inference to a missing data problem in which the researcher imputes missing counterfactual outcomes. This suggests matrix completion may prove equally versatile for evaluating policies such as COVID-19 mandates.

\begin{figure}
    \centering
    \includegraphics[width = .7\textwidth]{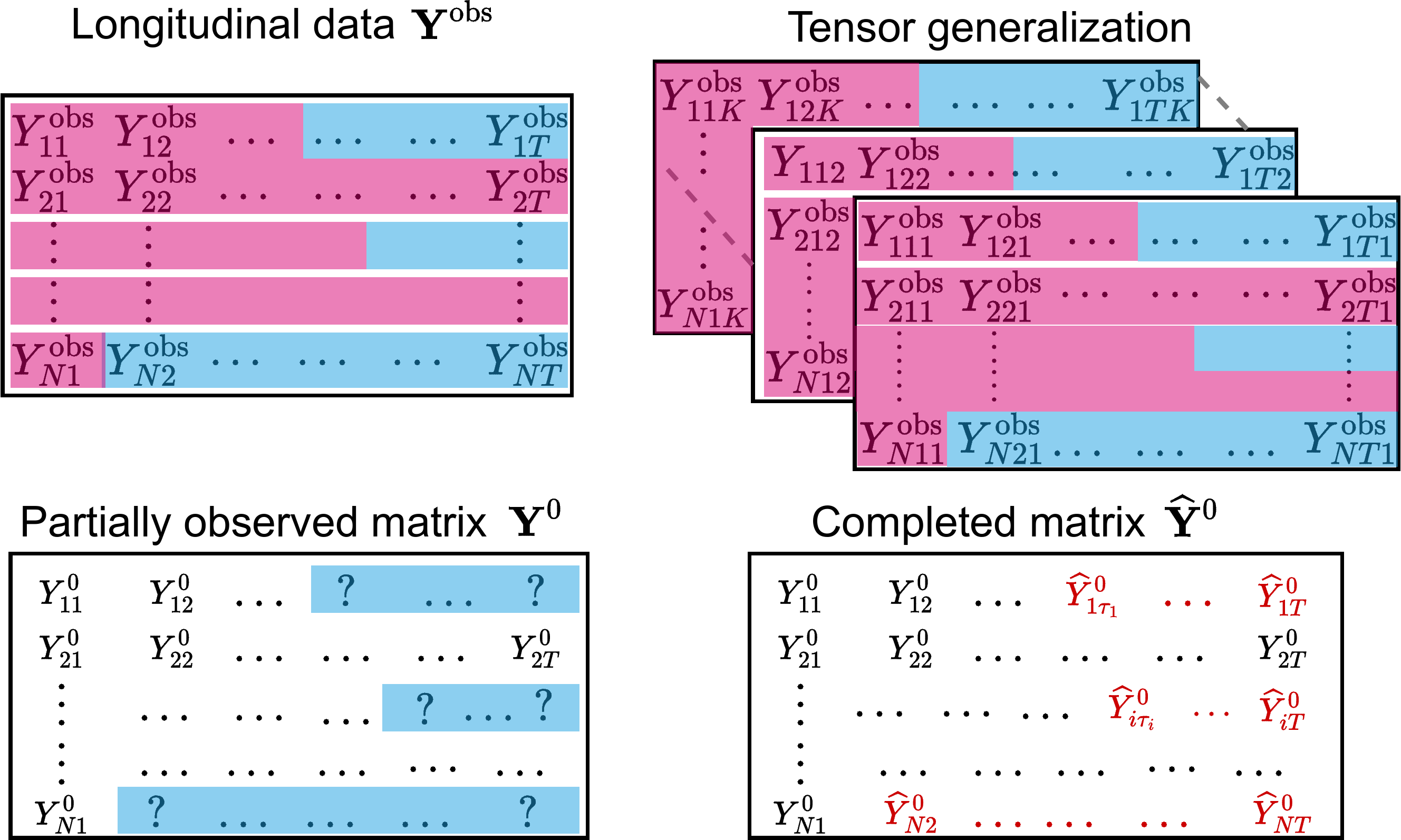}
    \vspace*{1ex}
    \caption{Illustration of the matrix completion approach to causal inference (top left) and the tensor generalization (top right). Counterclockwise from the top left: We observe a univariate longitudinal data matrix where units (rows) adopt treatment in different time periods (columns). Blue elements represent the outcome under treatment, and pink elements not under treatment. Bottom left: Of interest is $\M{Y}^0$, the matrix of outcomes that would have been observed had treatment never been adopted. $\M{Y}^0$ is only partially observed (incomplete) because the entries are missing for the units and time periods during which treatment was adopted. Bottom right: Completion of $\M{Y}^0$ allows estimation of causal estimands like the average treatment effect among the treated (ATT). Top right: tensor completion using a multivariate longitudinal data tensor can better impute the missing outcomes by borrowing information across outcomes. These additional outcomes improve the accuracy of imputations, but they are not otherwise of interest.}
\label{fig:matrix_tensor_completion}
\end{figure}

To fix ideas, consider the $N \times T$ matrix $\bf{Y}^{\text{obs}}$ depicted in the top left panel of Figure \ref{fig:matrix_tensor_completion}. Each row represents a location (\emph{e.g.} a U.S. state), each column represents a time period (\emph{e.g.} a season such as Spring 2020), and each entry denotes the outcome (\emph{e.g.} the number of fatalities on the log scale). Blue entries denote fatalities that occurred while the policy of interest was in place, and pink entries denote fatalities that occurred while the policy of interest was not in place.

One might wonder what $\bf{Y}^{\text{obs}}$ would have looked like had the policy never been in place and all entries were pink. We denote this hypothetical matrix $\bf{Y}^{\text{0}}$ and assume $\bf{Y}^{\text{0}}$ would agree with $\bf{Y}^{\text{obs}}$ on the pink entries but may disagree on the blue entries. Therefore, we proceed as though $\bf{Y}^{\text{0}}$ is partially observed, with incomplete entries obtained by discarding the blue entries of $\bf{Y}^{\text{obs}}$ as depicted in the bottom left panel of Figure \ref{fig:matrix_tensor_completion}. Matrix completion refers to the reconstruction of the missing entries using the observed entries to form the complete matrix, $\bf{\widehat Y}^{\text{0}}$, depicted in the bottom right panel. If the completed matrix reflects the outcome that would have been observed had the policy not been in place, any differences between the completed matrix, $\bf{\widehat Y}^{\text{0}}$, and the originally observed matrix, $\bf{Y}^{\text{obs}}$, can be attributed to the policy.

Matrix completion could refer to any algorithm that fills the entries of the incomplete matrix, $\bf{Y}^{\text{0}}$. In this paper, matrix completion refers to "low-rank" matrix completion, which assumes $\bf{Y}^{\text{0}}$ has approximately low rank. The two most popular approaches to low-rank matrix completion is low-rank factorization \cite{Koren2009, Ge2016} and regularized estimation, which enforces low rank structure for example by using a nuclear norm penalty \cite{Candes2009, Candes2010, Mazumder2010, Koltchinskii2011, Negahban2012}. Of these two methods, low-rank factorization is computationally simpler, but it suffers from the fact that the objective function is not convex and the rank of the underlying matrix is assumed to be known. Regularized estimation using the nuclear norm penalty avoids these drawbacks. Furthermore, the accuracy of the completed matrix is supported by rigorous statistical guarantees.

Tensor completion relies on the same principles as matrix completion, but it has a greater chance of recovering missing entries in the sense that the required fraction of observed entries can be significantly less than in the matrix case \cite{Barak2016, Montanari2018, Ghadermarzy2019, Xia2021}. Consider the $N \times T \times K$ tensor shown in the top right panel of Figure \ref{fig:matrix_tensor_completion}. The tensor can be thought of as a data cube storing data along three dimensions (\emph{e.g.}~unit, time, and outcome), or alternatively, as a sequence of matrices (\emph{e.g.}~unit and time) recorded along a single dimension (\emph{e.g.}~outcome). Each matrix can be completed separately using the observed entries only if there is enough information within each matrix to do so. But if the entries are related across matrices, the matrices may be able to borrow the information they need when it does not exist within.

Our work extends Athey \emph{et al.}~\cite{Athey2021}, which uses matrix completion to estimate causal effects with univariate longitudinal data. Athey \emph{et al.} extend the synthetic control approach of Abadie \emph{et al.}~\cite{Abadie2010} and explicitly connect matrix completion to the factor models that underlie synthetic controls. The extension is natural because factor models are a class of low-rank linear models for univariate longitudinal data, and matrix completion relies on a more general low-rank assumption. The extension to tensor completion that we propose in this paper allows for information sharing across matrices. To demonstrate this, we first consider a factor model in Subsection \ref{subsec:general_rationale}. We then consider tensors generally in Subsections \ref{subsec:computational_approach} and \ref{subsec:analysis}. Tensor completion is particularly useful when $N \gg \max\{T,K\}$. That is, the number of observations is much larger than the number of time periods or outcomes. We make this assumption throughout the paper as it characterizes the COVID-19 data discussed in Section \ref{sec:covid}.

\subsection{Combining multiple outcomes from a factor model}
\label{subsec:general_rationale}
Consider a factor model of the form 
\begin{equation*}
\E[\M{Y}] = \M{A} \M{B}^{\T} = \sum_{j = 1}^r \M{a}_j \M{b}_j^{\T}, \qquad \M{A} = [\M{a}_1 \; \ldots \; \M{a}_r], \quad  \M{B} = [\M{b}_1 \; \ldots \; \M{b}_r],
\end{equation*}
where $\M{Y}$ is an $N \times T$ random matrix, $\M{A}$ is an $N \times r$ matrix with column vectors $\M{a}_1, \; \ldots, \; \M{a}_r\text{,} \;$ $\M{B}$ is an $T \times r$ matrix with column vectors $\M{b}_1, \; \ldots, \; \M{b}_r\text{,} \;$ and $r$ denotes the number of factors. Note that this model includes two-way fixed effects as a special case: Define $\bm{1}_N$ to be the 1-vector of length $N$ and $\bm{1}_T$ the 1-vector of length $T$. Setting $\M{A} = [\bm{\alpha} \; \M{1}_N]$ and $\M{B} = [\bm{1}_T \; \bm{\beta}]$ yields the two-way fixed effects model with row effect $\bm\alpha$ and column effect $\bm\beta$.

The parameters $\M{A}$ and $\M{B}$ can be estimated from the data $\M{Y}$ without additional outcomes (under suitable identifying constraints), but we will show that under general conditions incorporating an additional $K-1$ random matrices, $\{ \M{Z}^{(k)} \}_{k = 1}^{K-1} = \M{Z}^{(1)}, \ldots, \M{Z}^{(K-1)}$, produces a more accurate estimator. For intuition, we consider the case where the outcome of interest and the additional outcomes are linearly dependent and no outcomes are missing. We will relax these assumptions in the following subsection. Linear dependence occurs under the model 
\begin{equation}\label{eq:model_control}
\E[\M{Z}^{(k)}] = \M{A} \bm{\Gamma}^{(k)} \M{B}^{\T}, \quad k=1,\ldots,K-1,     
\end{equation}    
where $\{ \M{\Gamma}^{(k)} \}_{k = 1}^{K-1} = \M{\Gamma}^{(1)}, \ldots, \M{\Gamma}^{(K-1)}$ is a set of $r$-by-$r$ matrices. The matrices $\M{Y}$ and $\{ \M{Z}^{(k)} \}_{k = 1}^{K-1}$ form an $N \times T \times K$ tensor, which we denote $\mathbfcal{Y} = (Y_{itk})$, where $Y_{it1} = Y_{it}$ and $Y_{itk} = Z^{(k-1)}_{it}$ when $k>1$, $1 \leq i \leq N$, $1 \leq t \leq T$. A Tucker-like decomposition of $\E[\mathbfcal{Y}]$  is given by
\begin{equation*}
\E[\mathbfcal{Y}] = \mathbfcal{G} \times_1 \M{A} \times_2 \M{B} \times_3 I_{K},    
\end{equation*}
where $\mathbfcal{G}$ is the $r$-by-$r$-by-$K$ tensor stacking the matrices $I_r$ and $\{ \bm{\Gamma}^{(k)} \}_{k = 1}^{K-1}$, and $I$ denotes an identity matrix whose dimension is specified in the subscript. See Appendix \ref{app:tensor} for a summary of relevant tensor linear algebra.

Estimating $\M{A}$, $\M{B}$, and $\{ \bm{\Gamma}^{(k)} \}_{k = 1}^{K-1}$ from $\mathbfcal{Y}$ amounts to estimating approximately $(N + T + K \cdot r) \cdot r$ parameters with a total of $N \cdot T \cdot K$ observations. The ratio of the number of parameters to the number of observations is thus
\begin{equation}\label{eq:df_ratio_general}
\frac{(N + T + K \cdot r) r}{N \cdot T \cdot K} = \frac{r}{T \cdot K} + \frac{r}{N \cdot K} + \frac{r^2}{N \cdot T} \lesssim 
\frac{r}{T \cdot K}   
\end{equation}
where we assume $N \gtrsim T \cdot K$ and $\lesssim, \gtrsim$ indicate upper and lower bounds up to absolute constants. In other words, we assume $N \cdot T$ and $N \cdot K$ is sufficiently large relative to $T \cdot K$ that the ratio of parameters to observations is approximately $r/(T\cdot K)$. By contrast, estimating $\M{A}$ and $\M{B}$ from the outcome of interest without the inclusion of additional outcomes results in a ratio of approximately $r / T$. It follows that multiple outcomes accommodate factor models with approximately $K$ times more factors than a single outcome model.

Linear dependence is unlikely to hold in practice. However, we show that the benefit of multiple outcomes extends to the case where the outcomes are not linearly dependent but the rank of the associated tensor increases slowly with the number of additional outcomes. We prove in Theorem \ref{theo:mse_bound} that the mean squared error (MSE) in estimating $\E[\mathbfcal{Y}]$ scales proportional to the right hand side of \eqref{eq:df_ratio_general}, whereas in the corresponding problem of estimating $\E[\M{Y}]$, the MSE scales proportional to $r / T$. This means that when few time periods are available, as is the case with COVID-19 data, only extremely low rank matrices can be recovered reliably without additional outcomes. Before discussing Theorem \ref{theo:mse_bound}, however, we briefly review  the proposed tensor completion approach.

\subsection{Combining multiple outcomes using tensor completion}
\label{subsec:computational_approach}
Let $\mathbfcal{Y} = (Y_{itk})$ denote the $N \times T \times K$ tensor obtained by stacking $\M{Y}$ and additional outcomes $\{ \M{Z}^{(k)} \}_{k = 1}^{K-1}$, and let $\mathcal{O} = \{(i,t,k): \; Y_{itk} \; \text{is observed} \}$ denote the set of observed indices. For example, $\M{Y}$ could be the partially observed matrix $\bf{Y}^{\text{0}}$, described at the beginning of this section.  Consider the associated Frobenius (semi)-norm with respect to $\mathcal{O}$ on the space $\R^{N \times T \times K}$ of the $N \times T \times K$ tensor: 
\begin{equation*}
\nnorm{\mathbfcal{A}}_{\mathcal{O}, \textsf{F}} = \left( \sum_{(i,t,k) \in \mathcal{O}} A_{itk}^2 \right)^{1/2}, \qquad \mathbfcal{A} = (A_{itk}).      
\end{equation*}

and the optimization problem 
\begin{equation}\label{eq:tensor_completion}
\min_{\mathbfcal{A} \in \R^{N \times T \times K}} \, \left\{ \nnorm{\mathbfcal{Y} - \mathbfcal{A}}_{\mathcal{O},\textsf{F}}^2 +  
\lambda \Omega(\mathbfcal{A}) \right\}, \quad \lambda > 0,
\end{equation}
where $\Omega: \R^{N \times T \times K} \rightarrow \R_+$ is a regularizer that encourages low-rank solutions. Unfortunately choosing $\Omega$ to be the tensor rank function makes problem \eqref{eq:tensor_completion} computationally intractable, as does convex surrogates such as the tensor nuclear norm (the analog to the nuclear norm for matrices)  \cite{Ghadermarzy2019, Hillar2013, Yuan2016}. A tractable approach is matricization (\emph{a.k.a.} matrix unfolding, see \ref{app:prelim}, \cite{Gandy2011, Tomioka2010} ), in which the regularizer $\Omega$ is the convex combination of matrix nuclear norms associated with each mode of the tensor. Matricization yields the convex optimization problem 
\begin{equation}\label{eq:nucnorm_tensor}
\min_{\mathbfcal{A} \in \R^{N \times T \times K}} \, \left\{ \nnorm{\mathbfcal{Y} - \mathbfcal{A}}_{\mathcal{O},\textsf{F}}^2 +
\lambda \sum_{j = 1}^{3} \alpha_j \nnorm{\mathbfcal{A}_{(j)}}_{*} \right\},
\end{equation}
where $\nnorm{\cdot}_{*}$ denotes the matrix nuclear norm, $\{\mathbfcal{A}_{(j)}\}_{j=1}^3$ denotes the modes of $\mathbfcal{A}$, and the $\{\alpha_j\}_{j = 1}^{3}$ are non-negative weights that sum to one. In our setting, the three modes are of dimension $N \times (T \cdot K)$, $T \times (N \cdot K)$, and $K \times (N \cdot T)$. Since we assume $N \gg \max\{T,K\}$ as discussed above, we consider only the first mode; the rank of the two other modes are bounded above by $T$ and $K$, respectively. Accordingly, we use the weights $\alpha_1 = 1$, $\alpha_2 = \alpha_3 = 0$, yielding the optimization problem
\begin{equation}\label{eq:nucnorm_matricization}
\min_{\Theta \in \R^{N \times (T \cdot K)}} \, \left\{ \nnorm{\mathbfcal{Y}_{(1)} - \Theta}_{\mathcal{O},\textsf{F}}^2 +
\lambda \nnorm{\Theta}_{*} \right\},
\end{equation}
where $\Theta = \mathbfcal{A}_{(1)}$.

We now show that when $N \gtrsim T \cdot K$, solving equation \eqref{eq:nucnorm_matricization} yields optimal rates up to logarithmic factors.

\subsection{Tensor completion improves the mean squared error bound relative to matrix completion} \label{subsec:analysis}
We present a non-asymptotic bound on the mean squared error of the estimator that minimizes \eqref{eq:nucnorm_matricization}. Our result follows established techniques from the literature on high-dimensional statistics as summarized in the monograph \cite{Wainwright2019}. We start by listing the underlying assumptions. 

{\bfseries (A1)}. The three-way tensor $\mathbfcal{Y}$ of dimensions $N$, $T$, and $K$ is generated according to the model 
\begin{equation*}
\mathbfcal{Y} = \mathbfcal{A}^* + \mathbfcal{E},    
\end{equation*}
where the tensor $\mathbfcal{E} = (\eps_{itk})$ contains independent random variables with zero mean satisfying the Bernstein moment condition $\max_{i,t,k} \E[\eps_{itk}^{2m}] \leq \frac{1}{2} (2m)! \beta^{2m - 2} \nu^2$ for all $m \geq 1$, for absolute constants $\beta, \nu > 0$.

{\bfseries (A2)}. Let $\mathcal{O}_1$ denote the set of observed entries in the matrix for the outcome of interest $\mathbf{Y}$, and let $\mathcal{O}_2, \ldots, \mathcal{O}_{K}$ denote the set of observed entries for the $K-1$ additional outcomes $\{ \mathbf{Z}^{(k)} \}_{k = 1}^{K-1}$. Let $n_{k} = |\mathcal{O}_k|$, $k = 1,\ldots,K$. We assume that the $\{ \mathcal{O}_{k }\}_{k = 1}^K$ are sampled independently and uniformly at random from $\{1,\ldots,N\} \times \{1,\ldots,T\}$. Moreover, we assume that
\begin{align*}
&n_1 \geq C \varrho_1^2 r \left[ (N + T) \log(N + T) \right] , \\[1ex]
&n_k \geq (1 - \overline{\alpha}) (N \cdot T) \vee C \varrho_{2:K}^2 r \left[ (N + T) \log(N + T) \right], \quad k = 2,\ldots,K, \\[-2.5ex]
\end{align*}
\begin{equation*}
\text{for} \; \, r = \text{rank}(\mathbfcal{A}_{(1)}^*), \quad \varrho_1 = \frac{\max_{i,t} |A_{it1}^*|}{\sqrt{\sum_{i,t} A_{it1}^{*2}} \, \big / \sqrt{N \cdot T}}, \quad \varrho_{2:K} = \frac{\max_{i,t,k\geq 2} |A_{itk}^*|}{\sqrt{\sum_{i,t, k \geq 2} A_{itk}^{*2}} \, \big / \sqrt{N \cdot T \cdot  (K-1)}},
\end{equation*}
where $C > 0$ and $0 \leq \overline{\alpha} < 1$.

{\bfseries (A3)} The dimensions of the tensor are sufficiently large in the sense that $N \vee (T \cdot K) \geq C_{\beta, \nu} \log\left[ N \vee (T \cdot K) \right]$.
\vskip1ex
Assumption {\bfseries (A1)} corresponds to an additive error model with i.i.d.~errors having sub-exponential tails, which facilitates the use of the Matrix Bernstein inequality. Assumption {\bfseries (A3)} requires a mild lower bound on the dimensions of the tensor. Assumption {\bfseries (A2)} specifies the sampling model for the observed entries in the tensor $\mathbfcal{Y}$: For the outcome of interest, we assume the number of observed entries is proportional to the underlying degrees of freedom (rank $r$ times the sum of the dimensions $N + T$), modulo a logarithmic factor. For each of the remaining outcomes, we assume the proportion observed is at least $\overline{\alpha}$. The required number of observed entries also depends on the ``spikiness" constants $\varrho_1$ and $\varrho_{2:K}$.

Assumption {\bfseries (A2)} or similar assumptions are commonly used in the matrix completion literature \cite{Koltchinskii2011, Negahban2012, Candes2009, Candes2010, Xia2021}. It can be relaxed at the expense of weaker recovery results or increased technical sophistication in the proofs. For example, the analysis in \cite{Athey2021} covers sampling models such as the staggered adoption setting in Figure \ref{fig:matrix_tensor_completion}. These extensions apply equally to our setting. However, since the primary purpose of our paper is to justify the use of multivariate outcomes in Section \ref{sec:covid}, we do not pursue these extensions in this paper.

\begin{theorem}\label{theo:mse_bound} Let $\widehat{\Theta}$ denote any minimizer of \eqref{eq:nucnorm_matricization} with corresponding back-folded tensor $\widehat{\Theta}^{\{ 1 \}}$. Under assumptions \emph{{\bfseries (A1)}} through \emph{{\bfseries (A3)}} for any $\lambda \geq 2\lambda_0$ with $\lambda_0 = 4 \nu \sqrt{2 C_1 \{ N \vee (T \cdot K) \}  \log\{2 \left[ N \vee (T \cdot K) \right]\}}$, it holds that
\begin{equation}\label{eq:mse_bound}
\frac{1}{\sqrt{N \cdot T \cdot K}} \nnorm{\mathbfcal{A}^* - \widehat{\Theta}^{\{1 \}}}_{\emph{\textsf{F}}} \leq 6 \lambda \sqrt{\frac{2r}{N \cdot T \cdot K}}.
\end{equation}
with probability at least 
$$ 1 - \frac{1}{2  \left[ N \vee ((T \cdot K) \right]} - 2 \exp(-(N + T) \log(N + T)) - 2 \exp\Big(-\big(N + (T \cdot (K-1)) \big) \, \log \big(N + (T \cdot (K-1)) \big) \Big),$$ 
where $C_1 = \frac{2}{1 - \overline{\alpha}}$ is a positive constant. 
\end{theorem}

We briefly discuss the meaning of Theorem 1; the proof is contained in the Appendix. The Theorem can be best unpacked by replacing $\lambda$ in \eqref{eq:mse_bound} by the ``optimal" choice $2 \lambda_0 \propto \sqrt{N \vee (T \cdot K) \log(N \vee (T \cdot K))}$, which yields the root mean square error bound 
\begin{equation*}
\frac{1}{\sqrt{N \cdot T \cdot K}} \nnorm{\mathbfcal{A}^* - \widehat{\Theta}^{\{1 \}}}_{\textsf{F}} \lesssim \sqrt{\frac{r}{N \wedge (T \cdot K)}} \cdot \sqrt{\log(N \vee (T \cdot K))},    
\end{equation*}
which is proportional to $\sqrt{r / (T \cdot K)}$ when $N \gtrsim T \cdot K$, modulo a logarithmic factor. The rate $\sqrt{r / (T \cdot K)}$ reflects the improvement that can be achieved by including additional outcomes relative to matrix completion (where $K = 1$), and the finding is in exact correspondence to \eqref{eq:df_ratio_general}, up to a logarithmic factor.

This result is easily extended to a model with covariates: 
\begin{equation}\label{eq:ext_to_covariate}
\E[\M{Y}] = \M{A}^* + \M{X}_N \M{B}^* \M{X}_T,
\end{equation}
where $\M{A}^* = (A_{it1}^*)$ is the top layer of an underlying low-rank tensor $\mathbfcal{A}^*$, $\M{X}_N \in \R^{N \times d_{N}}$ is a matrix of $d_{N}$ subject-specific covariates, $\M{X}_T \in \R^{d_{T} \times T}$ is a matrix of $d_{T}$ time-specific covariates, and the coefficient matrix $\M{B}^* \in \R^{d_{N} \times d_{T}}$ is an additional unknown parameter. Note that in the absence of time-specific covariates, we may simply use $\M{X}_T = I_{T}$ (and analogously $\M{X}_N = I_N$ for the absence of subject-specific covariates). The parameter $\M{B}^*$ can be estimated by a simple modification of the data fitting term in the objective \eqref{eq:nucnorm_matricization}; the resulting modified objective in $\Theta^*$ and $\M{B}^*$ remains (jointly) convex and can be solved easily with established methods in convex optimization. Under the assumption that $\min\{d_{N}, d_{T} \}$ is small, the matrix $\M{X}_N \M{B}^* \M{X}_T$ is of fixed (low) rank, which aligns with the low-rank/factor model assumption for $\E[\M{Y}]$ overall. Consequently, there is also no need in general to add a separate regularization term for the estimation of $\M{B}^*$. 

\subsection{Generalizing inference from matrix completion to tensor completion}
\label{subsec:inference}

We conclude this section by presenting three methods for constructing uncertainty intervals.  All three methods ---resampling, rotation-debiasing, and a full Bayesian approach---have been proposed for matrix completion (\emph{cf.} \cite{Athey2021}, \cite{chernozhukov2023inference}, \cite{zhai2023bayesian}), and we describe how they can be generalized to the tensor completion setting described in the previous subsections. Note that we consider uncertainty intervals for treatment effects, $\tau^*$, that can be written as a fixed linear combination of the entries of $\Theta^* = \mathbfcal{A}_{(1)}^*$. This includes the average effect of treatment on the treated (ATT), which we use in Section \ref{sec:covid}. 

Athey \emph{et al.} propose conducting inference via resampling after cross-validating the tuning parameter $\lambda$. In our generalization, we use leave-one-out cross-validation to identify the optimal value of $\lambda$. After selecting $\lambda$, we randomly permute the residuals of the observed entries associated with the outcome $\M Y$. The residuals for the additional outcomes are not permuted. Let $\widehat{\mathbfcal{E}}_{(1)}^{\pi}$ denote the permuted residuals matrix, $\pi = 1,\ldots, \Pi$. For each permutation $\pi$, we generate a new outcome $\widehat{\mathbfcal{Y}}_{(1)}^{\pi} = \widehat{\Theta} + \widehat{\mathbfcal{E}}_{(1)}^{\pi}$ and solve \eqref{eq:nucnorm_matricization} to obtain $\widehat{\Theta}_{\pi}$, from which we derive an estimate of $\tau^*$, $\widehat{\tau}_{\pi}$. The inner quantiles of $\{\widehat{\tau}_{\pi}\}_{\pi=1}^{\Pi}$ form an uncertainty interval for $\tau^*$.

\begin{comment}
\begin{algorithm}[H] 
\caption{Resampling Algorithm}
\label{alg:1}
\begin{algorithmic}[1]
\Statex
  \State {Select $\lambda$ by leave-one-out cross-validation.}
    \For{$\pi \text{ in } 1 \text{ to } \Pi$}     
    
        \State {$\widehat{\mathbfcal{E}}_{(1)}^{\pi}$ $\gets$ permute the residuals of the observed entries in the primary outcome $\M{Y}$,}

        \State {$\widehat{\mathbfcal{Y}}_{(1)}^{\pi}$ $\gets$ $\widehat{\Theta} + \widehat{\mathbfcal{E}}_{(1)}^{\pi}$,}

        \State {$\widehat{\Theta}_{\pi}$ $\gets$ solve \eqref{eq:nucnorm_matricization} using $\widehat{\mathbfcal{Y}}_{(1)}^{\pi}$,}

        \State {$\widehat{\Delta}_{\pi}$ $\gets$ $g(\widehat{\Theta}_{\pi})$.}
    \EndFor
    \State \Return {$\{\widehat{\Delta}_{\pi}\}_{\pi=1}^{\Pi}$}
\end{algorithmic}
\end{algorithm}
\end{comment}

Chernozhukov \emph{et al.} \cite{chernozhukov2023inference} propose a rotation-debiasing method that yields asymptotically normal estimators for linear combinations of rows of $\Theta^*$. In our generalization, we follow \cite{chernozhukov2019inference} and choose $\lambda\asymp \max\{ \sqrt{N}, \sqrt{T\cdot K}\}$. We then alternate between nuclear-norm penalized estimation and ordinary least squares regression: For a given row index $i \in \{1,\ldots,N\}$, we randomly partition $\{1,\ldots,N\} \setminus \{i\}$ into sets $\mathcal{I}$ and $\mathcal{I}^c$ with sizes $|\mathcal{I}|=N/2 -1$ and $|\mathcal{I}^c| = N/2$, and obtain intermediate estimates $\widetilde{\Theta}^{\mathcal{I}}$ and $\widetilde{\Theta}^{\mathcal{I}^c}$ of $\Theta^{*,\mathcal{I}}$ and $\Theta^{*,\mathcal{I}^c}$ using $\mathbfcal{Y}_{(1)}^{\mathcal{I}}$ and $\mathbfcal{Y}_{(1 )}^{\mathcal{I}^c}$, respectively, by solving a separate instance of \eqref{eq:nucnorm_matricization} each\footnote{In this paragraph, the superscripts $\mathcal{I}$ and $\mathcal{I}^c$ refer to the row-submatrices associated with these index sets}. We then obtain debiased estimates of the left and right singular vectors of $\Theta^{*,\mathcal{I}}$ and $\Theta^{*,\mathcal{I}^c}$ through ordinary least squares rotation-debiasing steps, resulting in the debiased estimator for the $i$-th row of $\Theta^*$ and the standard error. We use the estimates to construct the confidence interval $(\wh{\tau}_{i}^{\text{low}}, \wh{\tau}_{i}^{\text{up}})$ for $\tau_i^*$, which represents the contribution of row $i$ to the average treatment effect $\tau^*$. These steps are repeated for each $i$, spanning from $1$ to $N$; see Algorithm 2.1 in \cite{chernozhukov2023inference} for details. The debiased confidence interval for $\tau^*$ is obtained by averaging the sum of $\{ \wh{\tau}_{i}^{\text{low}} \}$ and $\{ \wh{\tau}_{i}^{\text{up}} \}$.

Zhai and Gutman \cite{zhai2023bayesian} propose a full Bayesian low-rank matrix completion algorithm, which modifies the Bayesian SVD model proposed by \cite{hoff2007model}. Our generalization follows the Gibbs Sampler described in Algorithm 3 \cite{zhai2023bayesian}, which assumes that the entries of the error matrix $\mathbfcal{E}_{(1)} = \mathbfcal{Y}_{(1)} - \Theta^*$ are independent normal random variables with mean zero and variance $1/\omega$ and assigns a gamma prior to $\omega$. We also put uniform priors on the left and right singular vectors of $\Theta^*$ and double exponential priors on the singular values, as well as standard normal priors on the coefficients of the covariates. We sample from the posterior distribution of $\Theta^*$, from which we approximate the distribution of $\tau^*$.

\section{A Reevaulation of COVID-19 Mandates} 
\label{sec:covid} 

We use the tensor completion approach outlined in Section \ref{sec:back} to estimate the effect of three different COVID-19 mandates: travel restrictions, mask-wearing directives, and vaccination requirements. Our application is motivated by a meta-analysis that failed to identify a single study of COVID-19 mask-wearing directives in which the risk of confounding was low \cite[see Figure 2]{talic2021effectiveness}. Concerns of confounding have been raised by a number of researchers, see for example \cite{zweig2021flawed}.

Our reevaluation of COVID-19 data uses additional control outcomes not otherwise of interest to stand in for confounding factors, a strategy discussed by \cite[Section 3.2]{rosenbaum1984association}, see also \cite[Chapter 21]{imbens2015causal}. As mentioned in Subsection \ref{subsec:general_rationale}, our work extends methods for causal inference with univariate longitudinal data proposed by Athey \emph{et al.}~\cite{Athey2021} and Abadie \emph{et al.}~\cite{Abadie2010}. Those methods use control units (\emph{e.g.} control jurisdictions) to adjust for confounding factors. They work by leveraging the staggered adoption of a policy to borrow from the experience of control jurisdictions, whose governments did not contemporaneously adopt the policy. The proposed tensor completion approach works by using control outcomes to increase the amount of information available for making fair comparisons between jurisdictions. This additional information can prove crucial when there are few time periods.

In Section \ref{sec:back}, we provide a theoretical argument for why the addition of control outcomes improves the accuracy of counterfactual imputations. In this Section, we demonstrate the importance of this accuracy by reevaluating the effect of COVID-19 mandates. We present the details of our analysis in three parts: In Subsection \ref{subsec:estimand}, we define the estimand; in Subsection \ref{subsec:data}, we summarize the data; and in Subsection
\ref{subsec:results}, we present the results of our reevaluation. Our main finding is that the models traditionally used to evaluate COVID-19 policies appear to overestimate the benefit of masking-wearing directives relative to travel restrictions. The proposed tensor completion approach suggests mask-wearing directives were not an effective alternative to travel restrictions. 

\subsection{We estimate the percent increase in fatalities that would have occurred had a mandate not been implemented}
\label{subsec:estimand}

Following the notation in Section \ref{sec:back}, let $i = 1, \ldots, N = 50$ denote the $i$th U.S. state and $t =1, \ldots, T = 8$ the $t$th season. Let $\M{Y}^{\text{obs}}$ denote the observed $N \times T$ matrix of COVID-19 fatalities. We consider two sets of state-season pairs, $\mathcal{O}$ and $\mathcal{M}$. The set $\mathcal{O}$ refers to the state-season pairs during which a given mandate was not in effect, and $\mathcal{M}$ refers to the state-season pairs for which the mandate was in effect.

We also define two additional $N\times T$ matrices of potential outcomes: Let matrix $\M{Y}^1$ have entries $Y_{it}^1$, denoting the number of COVID-19 fatalities that would occur were a given mandate adopted in state $i$ during season $t$, and let matrix $\M{Y}^0$ have entries $Y_{it}^0$, denoting the number of COVID-19 fatalities that would occur had the mandate not been adopted. We assume $\M{Y}^{\text{obs}}$ has entries $Y_{it}^0$ if $(i,t) \in \mathcal{O}$ and $Y_{it}^1$ if $(i,t) \in \mathcal{M}$

We are interested in the following estimand,  
\begin{equation*}
    \Delta = \frac{1}{|\mathcal{M}|}\sum_{(i,t)\in\mathcal{M}}\frac{Y_{it}^{0} - Y_{it}^{1}}{Y_{it}^{1}}.
\end{equation*}
$\Delta$ measures the average relative effect of treatment on the treated---the average percent increase in fatalities that would have occurred had states not implemented the given mandate (in the states and seasons for which the mandate was in fact implemented). Dividing by $Y^1_{it}$ accounts for the fact that states vary considerably in the observed number of fatalities; if we did not divide by $Y^1_{it}$ and computed the absolute effect, two states would dominate the average. 

The problem is that we only observe $\M{Y}^{\text{obs}}$, and the entries of $Y_{it}^{0}$ are missing when $(i,t) \in \mathcal{M}$---an impediment called the "fundamental problem of causal inference" \cite{holland1986statistics, rubin2005causal}. In the following subsections, we will compare methods, like tensor completion, that impute the missing entries of $\M Y^0$ from the non-missing entries so that $\Delta$ can be estimated,

\begin{equation*}
    \widehat{\Delta} = \frac{1}{|\mathcal{M}|}\sum_{(i,t)\in\mathcal{M}}\frac{\widehat Y_{it}^{0} - Y_{it}^{\text{obs}}}{Y_{it}^{\text{obs}}}.
\end{equation*}

We make the stable unit treatment value assumption (see \cite{imbens2015causal}): we assume there is only one version of a given mandate and the number of fatalities in any state and time period is unaffected by whether a mandate is adopted in a different state or time period. We believe this assumption is reasonable and consistent with the majority of longitudinal analysis of COVID-19 mandates as we discuss in Subsection \ref{subsec:data} after summarizing the data.

\subsection{We estimate the effect of mandates using eight seasons of COVID-19 fatalities across fifty U.S. states}
\label{subsec:data}

\begin{figure}
 \centering
  \includegraphics[scale = .2]{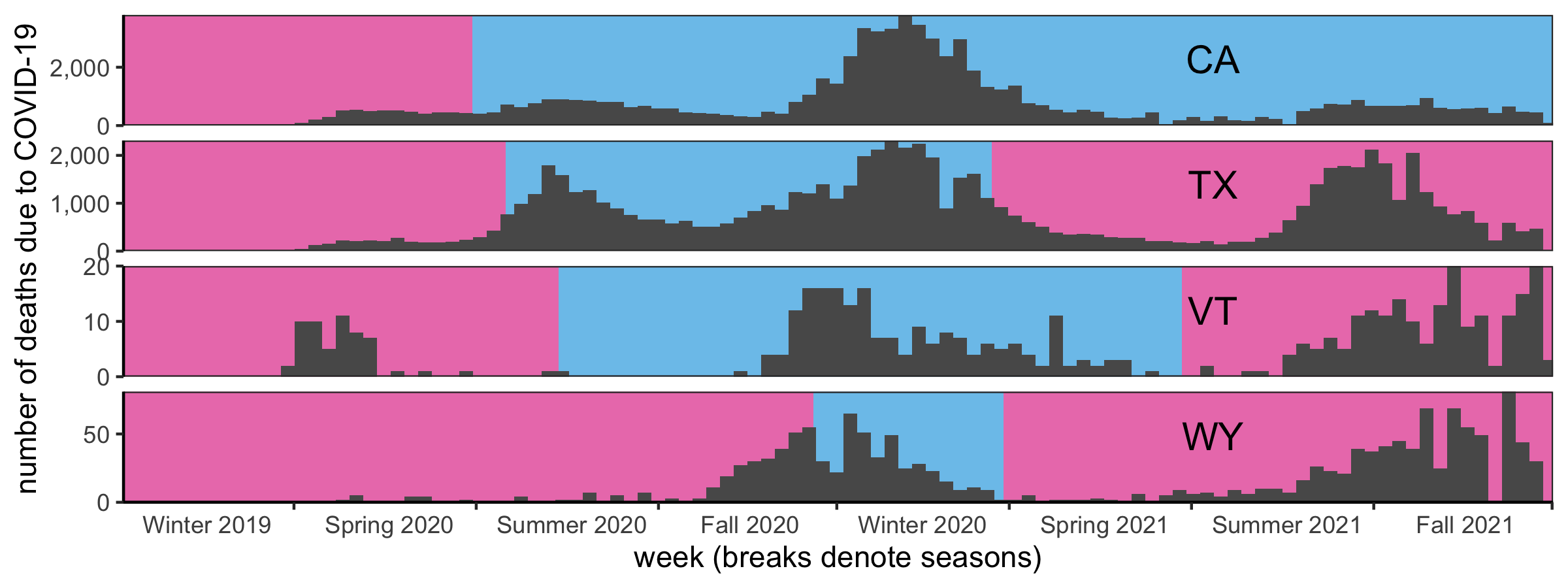}
  \caption{Weekly COVID-19 Deaths for the two most populous and two lease populous U.S. states. Colors represent whether a mask-wearing directive was in effect (pink - not in effect, blue - was in effect. Sources: OxCGRT \cite{hale2021global} and \href{ https://www.cdc.gov/coronavirus/2019-ncov/cases-updates/cases-in-us.html}{ CDC Case Task Force}}
  \label{fig:data_viz}
\end{figure}

The main outcome of interest is the number of COVID-19 fatalities reported by each U.S. state in the first two years of the pandemic as recorded by OxCGRT, the Oxford COVID-19 Government Response Tracker. Part of the data are visualized in four histograms in Figure \ref{fig:data_viz}. The top two histograms correspond with the two largest states (by population) and the bottom two histograms correspond with the two smallest. The dates are binned by week, and the ticks on the horizontal axis denote the seasons. The colors match Figure \ref{fig:matrix_tensor_completion}, representing whether a mandate was in effect: pink indicates that a mask-wearing directive was not in effect, and blue indicates that a mask-wearing directive was in effect.

We conduct our analysis at the season level for two reasons. The first reason is to reduce the impact of reporting errors. The existence of reporting errors is clear from Figure \ref{fig:data_viz}. One would expect the number of COVID-19 fatalities to change relatively slowly from week to week---the histogram of fatalities should be smooth. However, the data in Figure \ref{fig:data_viz} are somewhat jagged. On closer inspection, many fatalities at the end of the week or month are not reported until the beginning of the next week or month---certainly a result of the fatality reporting process and not COVID-19 itself. See \cite{kissane2021hownot} for details.

The second reason we aggregate fatalities by state and season is to reduce violations of the stable unit treatment assumption, mentioned in Section \ref{subsec:estimand}: We assume the mandate in any one state and period does not affect the number of fatalities in any other state and period. According to the \href{https://www.cdc.gov/coronavirus/2019-ncov/faq.html}{Center for Disease Control}, COVID-19 symptoms may appear two weeks after exposure, and infections can last a month or longer. It follows that the SUTVA assumption is likely violated at the day, week, or month level since a mandate in one month could reduce fatalities in a subsequent month without the mandate.

The second part of SUTVA is that there is only one version of treatment. In this paper, treatment is understood to be the intention of a state to influence behavior through legislation or executive order, such as by requiring masking. The mechanisms by which states chooses to influence behavior, such as the amount of enforcement or compliance, do not represent alternative treatments but rather the effect of the state's intention to influence behavior that we seek to measure. Studying the so-called intent to treat is important because policymakers typically choose which behavior to influence (\emph{e.g.} masking or travel) but cannot guarantee enforcement or compliance. However, the effect of a mask-wearing directive estimated in this paper should not be confused as the effect of wearing a mask.

\subsection{The proposed method fits the data better than five comparison methods and suggests mask-wearing directives were less effective}
\label{subsec:results}

We compare the proposed tensor completion approach to five other methods, which represent popular models used by researchers to study government mandates implemented during the pandemic. We find the proposed approach fits the data best as measured by mean squared error (estimated by leave-one-out cross validation, see Appendix for simulations). The proposed approach also indicates mask-wearing directives were between 2.5 and 7.5 times less effective than indicated by the comparison methods. In particular, the proposed method suggests mask-wearing directives prevented a minority of the fatalities that would have occurred had they not been implemented, whereas travel restrictions prevented the majority.

The first three comparison methods are log-linear models, the second two are matrix completion methods, and the last model is the proposed tensor completion method. The methods differ according to whether additional outcomes are (1) excluded, (2) included as covariates, or (3) included as additional outcomes. The three additional outcomes we consider in our analysis are the number of flu fatalities two years earlier (\emph{i.e.}~the seasons Winter 2017 to Fall 2019) as reported by the \href{https://wonder.cdc.gov/Deaths-by-Underlying-Cause.html}{Center for Disease Control}, and the average temperature and rainfall from all stations in the  \href{https://www.ncei.noaa.gov/products/land-based-station/global-historical-climatology-network-daily}{Global Historical Climatology Network} (\emph{i.e.}~the seasons Winter 2019 to Fall 2021). All models also include three "baseline" covariates: the log-number of vaccinated residents at the start of the season, the log-number of vaccinated residents at the end of the season, and the log-population (as determined by the 2020 decennial census) as an offset.\footnote{The only exception is that when we estimate the effect of vaccination requirements, we only include the log-population as an offset.}

The three log-linear models assume $Y_{it}^0$ follows a negative binomial distribution (NB) with mean $\text{exp}(\mu_{it})$ and size $\phi^{-1}$. In Log-linear Model (1), $\mu_{it}$ is decomposed into a state effect, $\theta_i$, a season effect, $\eta_t$, and a linear combination of covariates, $\sum_{p=1}^{P}\gamma_{p}X_{it}^{p} + \text{log} \; n_i$; Log-linear Model (2) augments Log-linear Model (1) by adding the additional outcomes as covariates on the log-scale; and Log-linear Model (3), augments Log-linear Model (1) by adding them as (conditionally) independent outcomes, offset by a factor $\delta^{k}_t$, which varies by season but not state:

\begin{align*}
& \text{{\bfseries Log-linear Model (1)}:} & Y_{it}^{0}  &\sim \text{NB}  ( \text{exp}(\mu_{it}), \; \phi  )  \\
&  & \mu_{it}  &= \theta_i + \eta_t + {\textstyle\sum_{p=1}^{P}\gamma_{p}X_{it}^{p}} + \text{log} \; n_i\\
& & & \\
& \text{{\bfseries Log-linear Model (2)}:} & Y_{it}^{0}  &\sim \text{NB}  ( \text{exp}(\mu_{it}), \; \phi  )  \\
&  & \mu_{it}  & = \theta_i + \eta_t + {\textstyle\sum_{p=1}^{P}\gamma_{p}X_{it}^{p}} +  {\textstyle\sum_{k=1}^{K-1}\gamma_{P+k} \text{log} \; Z_{it}^{(k)}} + \text{log} \; n_i \\
& & & \\
& \text{{\bfseries Log-linear Model (3)}:} & Y_{it}^{0} &\sim \text{NB} ( \text{exp}(\mu_{it} = \theta_i + \eta_t + {\textstyle\sum_{p=1}^{P}\gamma_{p}X_{it}^{p}} + \text{log} \; n_i), \; \phi )  \\
&  & Z_{it}^{(k)}   &\sim \text{NB} ( \text{exp}(\mu_{it} = \theta_i + \delta_t^k + {\textstyle\sum_{p=1}^{P}\gamma_{p}X_{it}^{p}} + \text{log} \; n_i), \; \phi ) \\
\end{align*}

We follow Section \ref{sec:back} in that the additional outcomes are denoted by the $N \times T$ matrix $\M{Z}^{(k)},$ $k \in \{1, \ldots, K - 1 = 3\}$ with entries $Z^{(k)}_{it}$, and the baseline covariates are denoted by the $N \times T$ matrix $\M{X}^p$, $p \in \{1, \ldots, P = 2\}$ with entries $X^p_{it}$. The population offset is denoted by $n_i$.

The parameters of all three log-linear models are estimated by maximum likelihood from the observed entries of $\M{Y}^0$. The missing entries of $\M{Y}^0$ are then imputed using the conditional expectation $\wh{Y}^0_{it} = \text{exp}(\wh \mu_{it})$, and $\wh{\Delta}$ is calculated as described in Subsection \ref{subsec:estimand}. Confidence interval are calculated with 99\% asymptotic coverage by deriving the multivariate normal approximation to the parameters of each model and applying the delta method.

The two matrix completion approaches and the proposed tensor completion approach were used to estimate $\Delta$ as described in Section \ref{sec:back}. These three approaches correspond loosely with the three log-linear models, but instead of maximizing the negative binomial likelihood, they solve the objective function \eqref{eq:nucnorm_matricization}, where the outcome is measured on the log-scale. In addition, the model Matrix Completion (1) includes baseline covariates; the second model, Matrix Completion (2), includes both baseline covariates and the additional outcomes as covariates; and the third model, Tensor Completion, adds baseline covariates but combines the outcomes as a tensor. Confidence intervals are calculated for the first two models using the resampling method, while confidence intervals for the third model are calculated with 99\% asymptotic coverage using the rotation-debiasing method, both described in Section \ref{subsec:inference}.\footnote{Rotation-debiasing was possible for Tensor Completion but not Matrix Completion (1) and Matrix Completion (2) because, without the inclusion of additional outcomes, the outcome matrix $\M Y^{\text{obs}}$ has too few columns. We believe the ability to calculate intervals via rotation-debiasing when there are few time periods is an additional benefit of the proposed approach.}

We apply all six methods to the data, yielding an estimate $\widehat{\Delta}$ for each of the three mandates: mask-wearing directives, travel restrictions, and vaccine requirements. A summary of the results are displayed in Table \ref{table:1}. Overall, we find that the proposed method (Tensor Completion) had the lowest mean squared error (MSE),  estimated by leave-one-out cross validation, where MSE is calculated for the outcome of interest only, as measured on the log-scale. (Improved accuracy is also demonstrated in simulation, see Appendix for details.) We also find that the five comparison methods estimate the effect of mask-wearing directives to be 2.5 to 7.5 times greater than that estimated by Tensor Completion. In contrast, Tensor Completion largely agrees with the comparison methods that travel restrictions had a treatment effect of around 2.3 (covered by all confidence intervals except one) and vaccine requirements had a treatment effect of around .79 (covered by all confidence intervals).

We interpret our findings as follows: The fact that the methods largely agree on the effect of travel restrictions and vaccine requirements provides greater evidence that the estimated effect is not due to unobserved confounding, while the fact that tensor completion disagrees with other estimates on the effect of mask-wearing directives provides evidence that these estimated effects are the result of unobserved confounding. In particular, most of the comparison methods suggest mask-wearing directives were an effective alternative to travel restrictions, while Tensor Completion suggests mask-wearing directives were not an effective alternative.

\begin{table}[ht]
\centering
\begin{scriptsize}
\begin{tabular}{lcccc cccc cccc}
  \toprule
      \multirow{2}{*}{Method} & \multicolumn{3}{c}{\textbf{Mask-Wearing Directives}} & &
      \multicolumn{3}{c}{\textbf{Travel Restrictions}} & &
      \multicolumn{3}{c}{\textbf{Vaccine Requirements}}
        \\[0.5ex]
      \cline{2-4} \cline{6-8} \cline{10-12}
      \specialrule{0em}{1pt}{1pt}
       & {$\Delta$} & {CI}   & MSE &  & {$\Delta$}
      &  {CI}  & MSE &  & {$\Delta$}
      &  {CI}  & MSE \\[0.5ex] 
  \hline 
  \specialrule{0em}{2pt}{2pt}
Log-linear Model (1) & $3.38$ & $(0.03, 6.72)$  & $0.64$ & & $2.34$  &  $(1.37, 3.30)$  & $0.59$  &  & $1.22$  & $(0.41, 2.03)$  & $0.63$ \\[0.5ex]
Log-linear Model (2) & $4.09$  & $(0.02, 8.16)$ & $0.62$  & & $2.25$  & $(1.34, 3.17)$  & $0.57$  & & $1.38$ & $(0.55, 2.22)$ & $0.63$ \\[0.5ex]

Log-linear Model (3) &  $2.10$ & $(1.32, 2.88)$  & $1.08$  &  & $9.83$    & $(7.44, 12.20)$  & $0.79$  & & $1.29$  & $(0.51, 2.06)$  & $1.10$ \\[0.5ex]

Matrix Completion (1) & $4.29$ & $(2.95, 7.49)$ &  $1.94$   & &   $1.70$ & $(1.39, 2.47)$  & $0.41$  & & $0.43$ & $(0.35, 0.79)$  & $0.40$ \\[0.5ex]

Matrix Completion (2) & $6.23$ & $(4.60, 10.41)$ & $2.10$   & & $2.54$    & $(1.94, 4.21)$  & $0.43$   & & $0.58$ & $(0.44, 0.95)$  & $0.47$ \\[0.5ex]

Tensor Completion & $0.83$ & $(0.42, 1.23)$  & $0.49$   & & $2.29$    & $(2.26,5.50)$  & $0.31$   & & $0.89$ & $(0.50, 1.26)$ & $0.33$ \\[0.5ex]
\bottomrule
\specialrule{0em}{3pt}{3pt}
\end{tabular}
\end{scriptsize}
\caption{Tensor Completion fits the data better than five comparison methods and suggests mask-wearing directives were less effective.}
\label{table:1}
\end{table}

We choose rotation-debiasing confidence intervals for Tensor Completion in Table \ref{table:1} because we believe the methodology is closest to the maximum likelihood-based intervals used for the log-linear models and thus allows for the fairest comparisons. Rotation-debiasing was not used for the matrix completion models because there were not enough time periods. We note that inference using the other approaches described in Subsection \ref{subsec:inference} yield different results---with resampling producing the smallest intervals and Bayesian SVD producing the largest. The intervals are summarized in Table \ref{table:2}. We do not believe one interval is necessarily preferable; each conveys a notion of uncertainty that could be of interest to policymakers.

\begin{table}[ht]
\centering
\begin{scriptsize}
\begin{tabular}{lcc cc cc}
  \toprule
      Method  & \textbf{Mask-Wearing Directives} & &
      \textbf{Travel Restrictions} & &
      \textbf{Vaccine Requirements} &\\[0.5ex]
  \hline 
    \specialrule{0em}{2pt}{2pt}
 Resample & $(0.74, 1.06)$  & & $(2.25, 2.33)$  &  &  $(0.81, 0.99)$
 \\[0.5ex]
 Rotation-Debiasing &  $(0.42, 1.23)$ & & $(2.26,5.50)$ & &  $(0.50, 1.26)$   \\[0.5ex]
  Bayesian SVD & $(0.47, 3.83)$ &  & $(3.07, 12.04)$ & &   $(0.06, 2.80)$ \\[0.5ex]
\bottomrule
\specialrule{0em}{3pt}{3pt}
\end{tabular}
\end{scriptsize}
\caption{Three different uncertainty intervals for the effect of mandates, estimated by Tensor Completion.}
\label{table:2}
\end{table}

\section{Discussion}
\label{sec:discussion}

We propose a new approach to causal inference with multivariate longitudinal data that uses tensor completion to impute missing counterfactual outcomes. Our motivation is to evaluate government mandates adopted during the pandemic such as travel restrictions, mask-wearing directives, and vaccine requirements. In Section \ref{sec:back}, we establish the general conditions under which the proposed method improves the accuracy of counterfactual imputations. We then show in Section \ref{sec:covid} that the proposed approach does in fact fit longitudinal data on COVID-19 fatalities better than traditional methods. We also show that the improved accuracy yields an important insight: mask-wearing directives were not the effective alternative to travel restrictions suggested by more traditional methods. We conclude that the proposed approach should be considered as an additional tool for studying policies such as government mandates, particularly when the number of time periods is small. In this section, we discuss several extensions that may be of interest to researchers considering this approach.

We estimate the effect of a mandate for the states and seasons during which the mandate was adopted (the so-called effect of treatment on the treated). This estimate is useful for discussing the amount of fatalities that were prevented by a mandate. However, policymakers may also be interested in the additional number of fatalities that could have been prevented had all states adopted the mandate in all seasons. The approach outlined in this paper can be used to estimate this so-called effect of treatment on the untreated, with the only modification being that a different set of observations would be considered missing.

In Section \ref{sec:covid}, the additional outcomes were chosen because their values did not depend on whether a state adopted the mandate. That is, $Z^{(k)}_{it}$ takes the same value regardless of whether $(i,t)\in\mathcal{O}$ or $(i,t)\in\mathcal{M}$. As a result, none of the entries of $\M{Z}^{(k)}$ are considered missing. However, the approach we outline in Section \ref{sec:back} is not limited to such outcomes. We could have included additional outcomes that were affected by the mandate, and these outcomes would be considered missing for the states and seasons during which the mandate was adopted. For example, we could include the number of flu deaths during the pandemic.

We assume in this paper that the additional outcomes are not of direct interest and are added in order to more accurately impute a single outcome of interest. However, in many cases the researcher is interested in more than one outcome. For example, policymakers may wish to choose mandates that reduce the number of COVID-19 fatalities, as well as the number of COVID-19 hospitalizations, the number of COVID-19 cases, and the number of COVID-19 positive tests. The approach we outline in Section \ref{sec:back} can accommodate multiple outcomes of interest, and these outcomes would be considered missing in the states and periods during which the mandate was adopted.

We conclude our discussion with a final thought on the inclusion of additional outcomes. Ultimately whether a variable should be treated as an outcome or as a covariate---or whether it should be excluded entirely---depends on the data generating process. Our work is based on the observation that the additional variables considered in Section \ref{sec:covid} should be treated as outcomes because they are influenced by similar unobserved factors (in the case of flu fatalities) or are surrogates for important unobserved factors, and thus should be considered as measured with error (in the case of average temperature and rain, which stand in for seasonal factors that influence human behavior). We believe the data support this observation: We find that by treating these variables as outcomes, the proposed method produces more accurate counterfactual imputations than traditional methods that treat these variables as covariates. This additional accuracy is important because governments increasingly rely on longitudinal data to choose among policies such as mandates during public health emergencies. Inaccurate methods can create the illusion that some policies are more effective than they actually are, leading governments to make suboptimal decisions that cost lives.

\section*{Acknowledgments}
This research was supported in part by a grant from the COVID-19 Database

%Bibliography
\bibliographystyle{unsrt}  
\bibliography{Arxiv.bib}  

\begin{thebibliography}{10}

\bibitem{holland1986statistics}
Paul~W Holland.
\newblock Statistics and causal inference.
\newblock {\em Journal of the American statistical Association},
  81(396):945--960, 1986.

\bibitem{rubin2005causal}
Donald~B Rubin.
\newblock Causal inference using potential outcomes: Design, modeling,
  decisions.
\newblock {\em Journal of the American Statistical Association},
  100(469):322--331, 2005.

\bibitem{willcox1938founder}
Walter~F Willcox.
\newblock The founder of statistics.
\newblock {\em Revue de l'Institut International de Statistique}, pages
  321--328, 1938.

\bibitem{graunt1939natural}
John Graunt.
\newblock {\em Natural and political observations made upon the bills of
  mortality}.
\newblock Johns Hopkins Press, 1939.

\bibitem{Rennie2005}
Jasson~DM Rennie and Nathan Srebro.
\newblock Fast maximum margin matrix factorization for collaborative
  prediction.
\newblock In {\em Proceedings of the 22nd international conference on Machine
  learning}, pages 713--719, 2005.

\bibitem{Koren2009}
Yehuda Koren, Robert Bell, and Chris Volinsky.
\newblock Matrix factorization techniques for recommender systems.
\newblock {\em Computer}, 42(8):30--37, 2009.

\bibitem{Candes2009}
Emmanuel~J Cand{\`e}s and Benjamin Recht.
\newblock Exact matrix completion via convex optimization.
\newblock {\em Foundations of Computational mathematics}, 9(6):717--772, 2009.

\bibitem{Candes2010}
Emmanuel~J Cand{\`e}s and Terence Tao.
\newblock The power of convex relaxation: Near-optimal matrix completion.
\newblock {\em IEEE Transactions on Information Theory}, 56(5):2053--2080,
  2010.

\bibitem{Koltchinskii2011}
Vladimir Koltchinskii, Karim Lounici, and Alexandre~B Tsybakov.
\newblock Nuclear-norm penalization and optimal rates for noisy low-rank matrix
  completion.
\newblock {\em The Annals of Statistics}, 39(5):2302--2329, 2011.

\bibitem{Negahban2012}
Sahand Negahban and Martin~J Wainwright.
\newblock Restricted strong convexity and weighted matrix completion: Optimal
  bounds with noise.
\newblock {\em Journal of Machine Learning Research}, 13(1):1665--1697, 2012.

\bibitem{Mazumder2010}
Rahul Mazumder, Trevor Hastie, and Robert Tibshirani.
\newblock Spectral regularization algorithms for learning large incomplete
  matrices.
\newblock {\em Journal of Machine Learning Research}, 11:2287--2322, 2010.

\bibitem{Ge2016}
R.~Ge, J.~Lee, and T.~Ma.
\newblock Matrix completion has no spurious local minimum.
\newblock In D.~Lee, M.~Sugiyama, U.~Luxburg, I.~Guyon, and R.~Garnett,
  editors, {\em {Advances in Neural Information Processing Systems}},
  volume~29, 2016.

\bibitem{Gao2016}
Chao Gao, Yu~Lu, Zongming Ma, and Harrison~H Zhou.
\newblock Optimal estimation and completion of matrices with biclustering
  structures.
\newblock {\em The Journal of Machine Learning Research}, 17(1):5602--5630,
  2016.

\bibitem{Li2020}
Tianxi Li, Elizaveta Levina, and Ji~Zhu.
\newblock Network cross-validation by edge sampling.
\newblock {\em Biometrika}, 107(2):257--276, 2020.

\bibitem{Gu2014}
Shuhang Gu, Lei Zhang, Wangmeng Zuo, and Xiangchu Feng.
\newblock Weighted nuclear norm minimization with application to image
  denoising.
\newblock In {\em Proceedings of the IEEE conference on computer vision and
  pattern recognition}, pages 2862--2869, 2014.

\bibitem{Natarajan2014}
Nagarajan Natarajan and Inderjit~S Dhillon.
\newblock Inductive matrix completion for predicting gene--disease
  associations.
\newblock {\em Bioinformatics}, 30(12):i60--i68, 2014.

\bibitem{Descary2019}
Marie-H{\'e}l{\`e}ne Descary and Victor~M Panaretos.
\newblock Functional data analysis by matrix completion.
\newblock {\em The Annals of Statistics}, 47(1):1--38, 2019.

\bibitem{Alidaee2020}
Hossein Alidaee, Eric Auerbach, and Michael~P Leung.
\newblock Recovering network structure from aggregated relational data using
  penalized regression.
\newblock {\em arXiv preprint arXiv:2001.06052}, 2020.

\bibitem{Barak2016}
B.~Barak and A.~Moitra.
\newblock Noisy tensor completion via the sum-of-squares hierarchy.
\newblock In {\em Conference on Learning Theory (COLT)}, pages 417--445, 2016.

\bibitem{Montanari2018}
A.~Montanari and N.~Sun.
\newblock Spectral algorithms for tensor completion.
\newblock {\em Communications on Pure and Applied Mathematics}, 71:2381--2425,
  2018.

\bibitem{Ghadermarzy2019}
N.~Ghadermarzy, Y.~Plan, and O.~Yilmaz.
\newblock Near-optimal sample complexity for convex tensor completion.
\newblock {\em Information and Inference: A Journal of the IMA}, 8:577--619,
  2019.

\bibitem{Xia2021}
D.~Xia, M.~Yuan, and C.-H. Zhang.
\newblock {Statistically optimal and computationally efficient low rank tensor
  completion from noisy entries}.
\newblock {\em The Annals of Statistics}, 49:76--99, 2021.

\bibitem{Athey2021}
Susan Athey, Mohsen Bayati, Nikolay Doudchenko, Guido Imbens, and Khashayar
  Khosravi.
\newblock {Matrix completion methods for causal panel data models}.
\newblock {\em Journal of the American Statistical Association}, pages 1--15,
  2021.

\bibitem{Abadie2010}
Alberto Abadie, Alexis Diamond, and Jens Hainmueller.
\newblock Synthetic control methods for comparative case studies: Estimating
  the effect of california’s tobacco control program.
\newblock {\em Journal of the American statistical Association},
  105(490):493--505, 2010.

\bibitem{Hillar2013}
C.~Hillar and L.-H. Lim.
\newblock Most tensor problems are np-hard.
\newblock {\em Journal of the ACM}, 60:1--39, 2013.

\bibitem{Yuan2016}
M.~Yuan and C.-H. Zhang.
\newblock On tensor completion via nuclear norm minimization.
\newblock {\em Foundations of Computational Mathematics}, 16:1031--1068, 2016.

\bibitem{Gandy2011}
Silvia Gandy, Benjamin Recht, and Isao Yamada.
\newblock Tensor completion and low-n-rank tensor recovery via convex
  optimization.
\newblock {\em Inverse problems}, 27(2):025010, 2011.

\bibitem{Tomioka2010}
Ryota Tomioka, Kohei Hayashi, and Hisashi Kashima.
\newblock {Estimation of low-rank tensors via convex optimization}.
\newblock {\em arXiv:1010.0789}, 2010.

\bibitem{Wainwright2019}
M.~Wainwright.
\newblock {\em High-dimensional statistics: A non-asymptotic viewpoint}.
\newblock Cambridge University Press, 2019.

\bibitem{chernozhukov2023inference}
Victor Chernozhukov, Christian Hansen, Yuan Liao, and Yinchu Zhu.
\newblock Inference for low-rank models.
\newblock {\em The Annals of statistics}, 51(3):1309--1330, 2023.

\bibitem{zhai2023bayesian}
Ruoshui Zhai and Roee Gutman.
\newblock A bayesian singular value decomposition procedure for missing data
  imputation.
\newblock {\em Journal of Computational and Graphical Statistics},
  32(2):470--482, 2023.

\bibitem{chernozhukov2019inference}
Victor Chernozhukov, Christian~Bailey Hansen, Yuan Liao, and Yinchu Zhu.
\newblock Inference for heterogeneous effects using low-rank estimations.
\newblock Technical report, CEMMAP working paper, 2019.

\bibitem{hoff2007model}
Peter~D Hoff.
\newblock Model averaging and dimension selection for the singular value
  decomposition.
\newblock {\em Journal of the American Statistical Association},
  102(478):674--685, 2007.

\bibitem{talic2021effectiveness}
Stella Talic, Shivangi Shah, Holly Wild, Danijela Gasevic, Ashika Maharaj,
  Zanfina Ademi, Xue Li, Wei Xu, Ines Mesa-Eguiagaray, Jasmin Rostron, et~al.
\newblock Effectiveness of public health measures in reducing the incidence of
  covid-19, sars-cov-2 transmission, and covid-19 mortality: systematic review
  and meta-analysis.
\newblock {\em bmj}, 375, 2021.

\bibitem{zweig2021flawed}
David Zweig.
\newblock The cdc’s flawed case for wearing masks in school.
\newblock {\em The Atlantic}, 2021.

\bibitem{rosenbaum1984association}
Paul~R Rosenbaum.
\newblock From association to causation in observational studies: The role of
  tests of strongly ignorable treatment assignment.
\newblock {\em Journal of the American Statistical Association},
  79(385):41--48, 1984.

\bibitem{imbens2015causal}
Guido~W Imbens and Donald~B Rubin.
\newblock {\em Causal inference in statistics, social, and biomedical
  sciences}.
\newblock Cambridge University Press, 2015.

\bibitem{hale2021global}
Thomas Hale, Noam Angrist, Rafael Goldszmidt, Beatriz Kira, Anna Petherick,
  Toby Phillips, Samuel Webster, Emily Cameron-Blake, Laura Hallas, Saptarshi
  Majumdar, et~al.
\newblock A global panel database of pandemic policies (oxford covid-19
  government response tracker).
\newblock {\em Nature Human Behaviour}, 5(4):529--538, 2021.

\bibitem{kissane2021hownot}
Erin Kissane and Jessica Malaty~Rivera.
\newblock How not to interpret covid-19 data.
\newblock {\em The COVID Tracking Project}, 2021.

\bibitem{Tropp2012}
J.A. Tropp.
\newblock User-friendly tail bounds for sums of random matrices.
\newblock {\em Foundations of Computational Mathematics}, 12(4):389--434, 2012.

\bibitem{Kolda2009}
Tamara~G Kolda and Brett~W Bader.
\newblock Tensor decompositions and applications.
\newblock {\em SIAM review}, 51(3):455--500, 2009.

\end{thebibliography}

\section*{Appendix}
\setcounter{section}{0}
\renewcommand{\thesection}{\Alph{section}}
\section{Proof of Theorem \ref{theo:mse_bound}}
Our proof can be decomposed into three main parts: I) Derivation of the main inequality, II) Empirical process control, 
III) Verification of a technical condition known as restricted strong convexity (\textsf{RSC}). These three parts are in
alignment with standard analysis schemes of matrix completion and related problems in high-dimensional estimation 
\cite{Wainwright2019}, while the details of the individual steps have been tailored to the setup under consideration. 

{\bfseries Part I: Derivation of the main inequality}\\
We start by expanding the model $\mathbfcal{Y} = \mathbfcal{A}^* + \mathbfcal{E}$ into its matricized 
counterpart: denote $\mathbfcal{A}_{(1)}^* = \Theta^*$, the observed (\emph{i.e.}~non-missing) entries are of the form 
\begin{equation*}
y_i = \Theta^*_{r(i), c(i)} + \xi_i,  \quad 1 \leq i \leq n,
\end{equation*}
where $n = \sum_{k = 1}^K n_k$ denotes the total number of observed entries including both the outcome of interest
and the $K-1$ control outcomes, $\{ r(i) \}_{i = 1}^n \subset \{1,\ldots,N\}$, $\{ c(i) \}_{i = 1}^n \subset \{1,\ldots,K\cdot T\}$ denote the row and column indices of the $n$ observed entries in the matricized 
tensor $\mathbfcal{Y}_{(1)}$, and the $\{ \xi_{i} \}_{i = 1}^n$ denote the corresponding random error terms
contained in $\mathbfcal{E}$. 

Accordingly, the objective function in \eqref{eq:nucnorm_matricization} can be expressed as 
\begin{equation*}
\sum_{i = 1}^n (y_i - \Theta_{r(i), c(i)})^2 + \lambda \nnorm{\Theta}_{\ast} =
\sum_{i = 1}^n (\Theta_{r(i), c(i)}^* - \Theta_{r(i), c(i)} + \xi_i)^2 + \lambda \nnorm{\Theta}_{\ast}. 
\end{equation*}
Since $\widehat{\Theta}$ is a minimizer of \eqref{eq:nucnorm_matricization}, we obtain the basic inequality
\begin{equation*}
\sum_{i = 1}^n (\Theta_{r(i), c(i)}^* - \widehat{\Theta}_{r(i), c(i)} + \xi_i)^2 + \lambda  \nnorm{\widehat{\Theta}}_{\ast} \leq \sum_{i = 1}^n \xi_i^2 + \lambda  \nnorm{\Theta^*}_{\ast}
\end{equation*}
Expanding the square on the left hand side and re-arranging terms yields 
\begin{equation}\label{eq:basic_inequality}
\nnorm{\mathscr{X}(\Theta^*) - \mathscr{X}(\widehat{\Theta})}_2^2 + \lambda  \nnorm{\widehat{\Theta}}_{\ast} \leq 2 \nscp{\widehat{\Theta} - \Theta^*}{\mathscr{X}^{\star}(\xi)} + \lambda  \nnorm{\Theta^*}_{\ast},  \qquad \xi = (\xi_i)_{i = 1}^n. 
\end{equation}
where $\mathscr{X}: \R^{N \times (T \cdot K)} \rightarrow \R^n$ denotes the linear operator 
defined by $\Theta \mapsto (\Theta_{r(i), c(i)})_{i = 1}^n$ and $\mathscr{X}^{\star}: \R^{n} \rightarrow \R^{N \times (T \cdot K)}$ denotes the adjoint operator given by $v \mapsto \sum_{i = 1}^n v_i e_{r(i)} f_{c(i)}^{\T}$ with
$\{ e_{i} \}_{i = 1}^N$ and $\{ f_{i} \}_{i = 1}^{T \cdot K}$ denoting the canonical basis vectors 
of $\R^N$ and $\R^{T \cdot K}$, respectively. 

Let the singular value decomposition of $\Theta^*$ be given by $\Theta^* = U \Sigma V^{\T}$ with 
$U \in \R^{N \times r}$, $V \in \R^{(T \cdot K) \times r}$, and $\Sigma \in \R^{r \times r}$, 
and consider the associated matrix subspace 
\begin{equation*}
\mathbb{M}^{\perp} = \{A \in \R^{N \times (T \cdot K)}:\;\, A = U_{\perp}  \Gamma  V_{\perp}^{\T}, \quad 
\Gamma \in \R^{(N - r) \times ((T \cdot K) -r)}\},
\end{equation*}
where the columns of $U_{\perp} \in \R^{N \times (N - r)}$ and $V_{\perp} \in \R^{(T \cdot K) -r}$
span the orthogonal complement of the column spaces of $U$ and $V$, respectively. Denote 
by $\mathbb{M}$ the orthogonal complement of $\mathbb{M}^{\perp}$. Then the following 
properties hold true (\emph{cf.}~\cite{Wainwright2019}, $\S$10.2.1). 
\begin{itemize}
\item[(S1)] Any element of $\mathbb{M}$ has rank at most $2r$, 
\item[(S2)] For any $\Theta \in \R^{N \times (T \cdot K)}$, we have $\nnorm{\Theta}_{*} = \nnorm{\Theta_{\mathbb{M}}}_{*} + \nnorm{\Theta_{\mathbb{M}^{\perp}}}_{*}$, where the superscripts 
$\mathbb{M}$ and $\mathbb{M}^{\perp}$ indicate the orthogonal projections of $\Theta$ onto these subspaces.  
\end{itemize}
Denote $\wh{\Delta} = \wh{\Theta} - \Theta^*$. It then  follows from \eqref{eq:basic_inequality} and 
the triangle inequality that
\begin{equation}\label{eq:cone_condtion_step1}
\lambda \nnorm{\wh{\Delta}_{\mathbb{M}^{\perp}}}_{*} \leq 2 \nscp{\wh{\Delta}}{\mathscr{X}^{\star}(\xi)} + \lambda \nnorm{\wh{\Delta}_{\mathbb{M}}}_{*},
\end{equation}
where we have used that $\Theta^*_{\mathbb{M}^{\perp}} = 0$. The Von Neumann trace inequality yields 
\begin{equation*}
\lambda \nnorm{\wh{\Delta}_{\mathbb{M}^{\perp}}}_{*} \leq \nnorm{\wh{\Delta}}_{*} \nnorm{2 \mathscr{X}^{\star}(\xi)}_{\text{op}} + \lambda \nnorm{\wh{\Delta}_{\mathbb{M}}}_{*},
\end{equation*}
where $\nnorm{\cdot}_{\text{op}}$ denotes the operator norm. As a result, conditional on the event 
\begin{equation}\label{eq:emp_process}
\textsf{E} = \{ \nnorm{2 \mathscr{X}^{\star}(\xi)}_{\text{op}} \leq \lambda_0 \},
\end{equation}
we have for any $\lambda \geq 2 \lambda_0$
{\small \begin{equation*}
\wh{\Delta} \in \mathbb{C}(\mathbb{M}; \lambda, \lambda_0) \coloneq \left \{ \Delta \in \R^{N \times (T \cdot K)}: \; 
\nnorm{\Delta_{\mathbb{M}^{\perp}}}_{*} \leq \frac{\lambda + \lambda_0}{\lambda - \lambda_0}  \nnorm{\Delta_{\mathbb{M}}}_* \right \}  \subseteq    \left\{ \Delta \in \R^{N \times (T \cdot K)}: \; 
\nnorm{\Delta_{\mathbb{M}^{\perp}}}_{*} \leq 3 \nnorm{\Delta_{\mathbb{M}}}_* \right \}
\end{equation*}}
Consequently, conditional on \textsf{E}, with the same reasoning that led to \eqref{eq:cone_condtion_step1}, we obtain from \eqref{eq:basic_inequality}
\begin{align}
\nnorm{\mathscr{X}(\wh{\Delta})}_2^2 = \nnorm{\mathscr{X}(\Theta^*) - \mathscr{X}(\widehat{\Theta})}_2^2 &\leq \lambda_0 \nnorm{\wh{\Delta}}_{*} + \lambda   \nnorm{\wh{\Delta}_{\mathbb{M}}}_{*} \notag \\
&=  \lambda_0 (\nnorm{\wh{\Delta}_{\mathbb{M}}}_{*}  + \nnorm{\wh{\Delta}_{\mathbb{M}^{\perp}}}_{*} ) + \lambda   \nnorm{\wh{\Delta}_{\mathbb{M}}}_{*} \notag \\
&\leq 3 \lambda \nnorm{\wh{\Delta}_{\mathbb{M}}}_{*} \leq 3 \sqrt{2r} \lambda \nnorm{\wh{\Delta}}_{\textsf{F}} \label{eq:fromcone},
\end{align}
where we have used property (S1) above in the last inequality. 

Conditional on the event 
\begin{equation}\label{eq:ev_RSC}
\textsf{RSC} = \{ \nnorm{\mathscr{X}(\wh{\Delta})}_2^2  > \kappa \nnorm{\wh{\Delta}}_{\textsf{F}}^2 \}    
\end{equation} 
for some constant $\kappa > 0$, we then obtain the following bound from \eqref{eq:fromcone}: 
\begin{equation*}
\frac{\nnorm{\wh{\Delta}}_{\textsf{F}}}{\sqrt{N \cdot T \cdot K}} \leq \frac{3 \sqrt{2}}{\kappa} \lambda \sqrt{\frac{r}{N \cdot T \cdot K}}. 
\end{equation*}
In order to conclude the proof of the theorem, we elaborate on the events \eqref{eq:emp_process} and 
\eqref{eq:ev_RSC} in part II and part III, respectively. 

{\bfseries Part II: Empirical process control}\\
In order to obtain a suitable bound $\lambda_0$ in \eqref{eq:emp_process}, we shall invoke assumption 
{\bfseries (A1)} and the matrix Bernstein inequality \cite{Tropp2012}. 

We have 
\begin{equation}\label{eq:decomposition_empiricalprocess}
\mathscr{X}^{\star}(\xi) = \sum_{i = 1}^n R_i, \qquad R_i \coloneq \xi_i e_{r(i)} e_{c(i)}^{\T}.
\end{equation}
for i.i.d.~random matrices $\{ R_i \}_{i = 1}^n$. Consider the corresponding symmetric matrices 
\begin{equation}\label{eq:Qi}
Q_i = \begin{bmatrix}
                                    0   & \xi_i e_{r(i)} e_{c(i)}^{\T} \\
                                    \xi_i e_{c(i)} e_{r(i)}^{\T}  &  0 
                                    \end{bmatrix}, \qquad 1 \leq i \leq n. 
\end{equation}
Observe that $\E[Q_i^m] = 0$ for all odd integers $m \geq 1$, $1 \leq i \leq n$. Moreover, 
\begin{equation*}
\E[Q_i^{2m}] = \E[\xi_i^{2m}]     \begin{bmatrix}
                                    \E[e_{r(i)} e_{r(i)}^{\T}] &  0\\
                                    0 &  \E[e_{c(i)} e_{c(i)}^{\T}]
                                    \end{bmatrix}, \quad 1 \leq i \leq n. 
\end{equation*}    
Now note that since the $\{ \xi_i \}_{i = 1}^n$ satisfy the Bernstein condition specified
in {\bfseries (A1)}, the following property holds:
\begin{equation}\label{eq:Bernstein_matrix_property}
\E[Q_i^{2m}] \preceq \frac{1}{2} (2m!) \beta^{2m - 2} \underbrace{\nu^2   \begin{bmatrix}
                                    \E[e_{r(i)} e_{r(i)}^{\T}] &  0\\
                                    0 &  \E[e_{c(i)} e_{c(i)}^{\T}]
                                    \end{bmatrix}}_{=\E[Q_i^2]}, \qquad \nu^2 = \E[\xi_i^2], \quad 1 \leq i \leq n, 
\end{equation}    
where $\preceq$ denotes positive semidefinite ordering, \emph{i.e.}~$A \preceq B$ for symmetric matrices $A$ and $B$ iff $B - A$ is positive semidefinite.  

In light of property \eqref{eq:Bernstein_matrix_property}, we are in position to apply the matrix Bernstein inequality in the form of Theorem 6.17 in \cite{Wainwright2019}\footnote{Strictly speaking, the matrix Bernstein inequality requires the distribution of the noise 
$\{ \xi \}_{i = 1}^n$ to be symmetric around the origin; non-symmetric distributions can be accommodate at the expense of slightly worse constants using the symmetrization argument detailed in \cite{Wainwright2019}. For brevity, we do not reproduce this argument here.}, which states that for any $\delta \geq 0$
\begin{equation}\label{eq:matrix_bernstein}
\P \left(\norm{\sum_{i = 1}^n Q_i}_{\text{op}} \geq \delta \right) \leq 2 \cdot \text{rank}\left(\sum_{i = 1}^n \E[Q_i^2] \right) \exp\left(-\frac{\delta^2}{2(\sigma^2 + \beta \delta)} \right), \qquad \sigma^2 \coloneq \norm{\sum_{i = 1}^n \E[Q_i^2]}_{\text{op}},
\end{equation}
and $\beta$ as in \eqref{eq:Bernstein_matrix_property}. In the sequel, we shall derive an upper bound on 
$\sigma^2$. 

By assumption {\bfseries (A2)}, we have 
\begin{equation*}
  \begin{bmatrix} \E[e_{r(i)} e_{r(i)}^{\T}] &  0\\
                                    0 &  \E[e_{c(i)} e_{c(i)}^{\T}]  \end{bmatrix} = \begin{bmatrix}
                               \frac{1}{N} I_N&   & &  &\\  
                               &  \frac{n_1}{\sum_{k = 1}^K n_k} \frac{1}{T} I_T & & & \\
                               & & \frac{n_2}{\sum_{k = 1}^K n_k} \frac{1}{T} I_T & & \\
                               & & & \ddots & \\ 
                                 & & & & \frac{n_K}{\sum_{k = 1}^K n_k} \frac{1}{T} I_T 
                              \end{bmatrix}
\end{equation*}
Moreover, note that again by assumption {\bfseries (A2)}, $n_k \big / \sum_{k = 1}^K n_k \leq \frac{1}{(1  - \overline{\alpha}) (K-1)}$, $k = 1,\ldots,K$. It follows that 
\begin{equation*}
  \norm{\begin{bmatrix} \E[e_{r(i)} e_{r(i)}^{\T}] &  0\\
                                    0 &  \E[e_{c(i)} e_{c(i)}^{\T}]  \end{bmatrix}}_{\text{op}} \leq C_1 \left\{\frac{1}{N} \vee \frac{1}{T \cdot K}\right\}, \quad C_1 \coloneq \frac{1}{1 - \overline{\alpha}}\frac{K}{K-1} \leq \frac{2}{1 - \overline{\alpha}}
\end{equation*}    
This implies that the quantity $\sigma^2$ in \eqref{eq:Bernstein_matrix_property} can be bounded as
$\sigma^2 \leq \overline{\sigma^2} \coloneq C_2 \{ N \vee (T \cdot K) \}$ with $C_2 = \nu^2 C_1$. Invoking \eqref{eq:Bernstein_matrix_property} with the choice 
\begin{equation*}
\delta = \sqrt{8 \overline{\sigma^2} \log\{2 \left[ N \vee (T \cdot K) \right]\} } 
\end{equation*}
and using the assumption {\bfseries (A3)} that 
\begin{equation*}
N \vee (T \cdot K) \geq 8 \left(\frac{\beta}{\nu^2} \right)^2 \log\{2 \left[ N \vee (T \cdot K) \right]\}
\end{equation*}
(which implies that $\beta \delta \leq \overline{\sigma^2}$), we obtain after collecting all terms
\begin{equation}\label{eq:use_of_bernstein_final}
\P \left(\norm{\sum_{i = 1}^n Q_i}_{\text{op}} \geq \nu \sqrt{8 C_1 \{ N \vee (T \cdot K) \}  \log\{2 \left[ N \vee (T \cdot K) \right]\}} \right)  \leq \frac{1}{2  \left[ N \vee (T \cdot K) \right]}.   
\end{equation}    
Combining \eqref{eq:decomposition_empiricalprocess}, \eqref{eq:Qi} and \eqref{eq:use_of_bernstein_final} yields
that the event \textsf{E} in \eqref{eq:emp_process} holds with probability at least $1 - \frac{1}{2  \left[ N \vee (T \cdot K) \right]}$ with the choice $\lambda_0 = 4 \nu \sqrt{2 C_1 \{ N \vee (T \cdot K) \}  \log\{2 \left[ N \vee (T \cdot K) \right]\}}$. 

{\bfseries Part III: Verification of restricted strong convexity}\\
We finally turn to the restricted strong convexity condition specified in event \textsf{RSC} in \eqref{eq:ev_RSC}.
Let us structure $\wh{\Delta} = [\wh{\Delta}_1 \; \wh{\Delta}_{2:K}]$, where $\wh{\Delta}_1$ and $\wh{\Delta}_{2:K}$
denote the column submatrices of $\wh{\Delta}$ containing the first $T$ and the remaining $T \cdot (K-1)$ columns, respectively.  
and accordingly define operators $\mathscr{X}_1 : \R^{N \times T} \rightarrow \R^{n_1}$ and $\mathscr{X}_{2:K} : \R^{N \times (T \cdot (K-1))} \rightarrow \R^{n - n_1}$ extracting
entries corresponding to $\mathcal{O}_1$,  and $\cup_{k = 2}^K \mathcal{O}_k$. We then have 
\begin{equation*}
\mathscr{X}(\wh{\Delta}) = \left[ \begin{array}{c}
     \mathscr{X}_1(\wh{\Delta}_1) \\
       \mathscr{X}_{2:K}(\wh{\Delta}_{2:K})
\end{array}  \right],
\end{equation*}
as well as 
\begin{equation*}
\nnorm{\mathscr{X}(\wh{\Delta})}_2^2 = \nnorm{\mathscr{X}_1(\wh{\Delta}_1)}_2^2  +  \nnorm{\mathscr{X}_{2:K}(\wh{\Delta}_{2:K})}_2^2 .    
\end{equation*}
Consider the events
\begin{equation}\label{eq:RSC_0}
\textsf{RSC}_1 = \{ \nnorm{\mathscr{X}_{1}(\wh{\Delta}_1)}_2^2 \geq \kappa \nnorm{\wh{\Delta}_1}_2^2 \}, \qquad 
\textsf{RSC}_{2:K} =  \{ \nnorm{\mathscr{X}_{2:K}(\wh{\Delta}_{2:K})}_2^2 \geq \kappa \nnorm{\wh{\Delta}_{2:K}}_2^2 \}. 
\end{equation}
and observe that $\textsf{RSC}_1 \cap \textsf{RSC}_{2:K}$ implies \textsf{RSC}, hence it suffices to show that the intersection of these two events occurs with the stated probability. 

Using Assumption {\bfseries (A2)}, Theorem 10.17 in \cite{Wainwright2019} implies
that the events in \eqref{eq:RSC_0} hold with $\kappa = 1/2$ with probability at least  $1 - 2 \exp(-(N + T) \log(N + T))$ and $1 - 2 \exp(-(N + (T \cdot (K-1))) \log(N + (T \cdot (K-1))))$, respectively. This concludes the proof of the theorem. 
%Since the error bound to be derived will be proportional to the scale of the error terms, observe that bounding the %MSE
%\begin{equation*}
%\frac{1}{N \cdot T \cdot (K+1)} \nnorm{\widehat{\Theta} - \Theta^*}_{\textsf{F}}^2    
%\end{equation*}    
%is equivalent to bouding the squared Frobenius norm $\nnorm{\widehat{\Theta} - \Theta^*}_{\textsf{F}}^2$ 
%under the observation model
%\begin{equation*}
%\widetilde{y}_i = \Theta^*_{r(i), c(i)} + \frac{\xi_i}{\sqrt{N \cdot T \cdot (K+1)}},  \quad 1 \leq i \leq n,
%\end{equation*}
%which in turn is equivalent to the observation model 
%\begin{equation*}
%\widetilde{\widetilde{y}}_i = \sqrt{N \cdot T \cdot (K + 1)} \Theta^*_{r(i), c(i)} +\xi_i ,  \quad 1 \leq i \leq n,
%\end{equation*}
%and denote the 
%entries of the matrix $\mathbfcal{E}_{(1)} $
\section{Tensor Algebra}\label{app:tensor}
In order to keep this paper self-contained, we here review selected notions concerning computations with 
tensors (of order three, for simplicity). For a more detailed account, we refer to the excellent survey \cite{Kolda2009}. 

\subsection{Preliminaries}\label{app:prelim}
A real-valued tensor of order three with dimensions $(d_1, d_2, d_3) \in \mathbb{N}^3$ is given by $\mathbfcal{T} = (T_{ijk})$ with $T_{ijk} \in \R$, 
$1 \leq i \leq d_1$, $1 \leq j \leq d_2$, and $1 \leq k \leq d_3$. 

The $\ell$-th mode (or $\ell$-th unfolding) of $\mathbfcal{T}$, $1 \leq \ell \leq 3$, is denoted by 
$\mathbfcal{T}_{(\ell)}$, a matrix of dimension $d_{\ell} \times \prod_{m=1, m \neq \ell}^3 d_m$ whose entries are given by 
\begin{align*}
&(\mathbfcal{T}_{(1)})_{qs} = T_{q j(s) k(s)}, \quad j(s) = s  \div  d_2, \; k(s) = \lceil s / d_2 \rceil, \quad q=1,\ldots,d_1, \; s=1,\ldots,d_2 \cdot d_3,  \\
&(\mathbfcal{T}_{(2)})_{qs} = T_{i(s) q k(s)}, \quad i(s) = s \div d_1, \; k(s) = \lceil s/d_1 \rceil, \quad q=1,\ldots,d_2, \; s=1,\ldots,d_1 \cdot d_3, \\
& (\mathbfcal{T}_{(3)})_{qs} = T_{i(s) j(s) q}, \quad i(s) = s \div d_1, \; k(s) = \lceil s / d_1\rceil, \quad q=1,\ldots,d_3, \; s=1,\ldots,d_1 \cdot d_2,
\end{align*}
where $\lceil \cdot \rceil$ denotes the ceiling operation and $\div$ denotes modulo division, with the modification that zero
rest is replaced by the divisor, \emph{i.e.}~for any integer $k \geq 1$, we have $(k \cdot d_1) \div d_1 = d_1$. 

The reverse operation to unfolding is referred to as back-folding: specifically, for a matrix $\M{M}$ of dimension $d_{\ell} \times \prod_{m=1, m \neq \ell}^3 d_m$, we define a tensor $\mathbf{M}^{\{\ell\}}$ of dimensions $(d_1, d_2, d_3)$ by 
$(\M{M}^{\{\ell \}})_{(\ell)} = \M{M}$, $\ell=1,\ldots,3$.

\subsection{CP decomposition and tensor rank}\label{app:CP_rank}
Consider vectors $a^{(\ell)} \in \R^{d_{\ell}}$, $1 \leq \ell \leq 3$. The outer product of these vectors is expressed
as 
\begin{equation*}
a^{(1)} \circ a^{(2)} \circ a^{(3)} 
\end{equation*}
which yields a tensor of dimension $(d_1, d_2, d_3)$ with entries $T_{i j k} = a_i^{(1)} \cdot a_j^{(2)} \cdot a_k^{(3)}$,
$1 \leq i \leq d_1$, $1 \leq j \leq d_2$, $1 \leq k \leq d_3$ (note that this generalizes the usual outer product 
of two vectors $u$ and $v$ given by $u v^{\T} = (u_i v_j)$). 

The CP decomposition decomposes a tensor into a sum of outer products (with a minimum number of summands), \emph{i.e.} 
\begin{equation*}
\mathbfcal{T} = \sum_{m = 1}^r a_m^{(1)} \circ a_m^{(2)} \circ a_m^{(3)},     
\end{equation*}
where $r$ is called the (CP)-rank (or simply rank) of the tensor $\mathbfcal{T}$. 

\subsection{\texorpdfstring{$\ell$}{l}-mode product and Tucker decomposition}\label{app:tucker}
The $\ell$-mode product of a tensor $\mathbfcal{T}$ of dimensions $(d_1, d_2, d_3)$ with matrix $\M{M}$ of dimension
$m \times d_{\ell}$, denoted by $\mathbfcal{T}\times_{\ell} \M{M}$, yields a tensor $\mathbfcal{U}$ of dimension
\begin{equation*}
\begin{cases}
(m, d_2, d_3),   \quad \ell = 1,\\
(d_1, m, d_3),   \quad \ell = 2,\\
(d_1, d_2, m), \quad \ell =3,
\end{cases}
\end{equation*}
such that its $\ell$-th mode is given by the matrix $\mathbfcal{U}_{(\ell)} = \M{M} \mathbfcal{T}_{(\ell)}$, $\ell=1,2,3$. 

The Tucker decomposition decomposes the tensor $\mathbfcal{T}$ as 
\begin{equation*}
\mathbfcal{T} = \mathbfcal{G} \times_1 \M{A}^{(1)} \times_2 \M{A}^{(2)} \times_3 \M{A}^{(3)},     
\end{equation*}
where the $\M{A}^{(\ell)}$ are (usually orthonormal) matrices of dimension $d_{\ell} \times m_{\ell}$, $\ell =1,\ldots,3$, and 
$\mathbfcal{G}$ is a tensor of dimension $(m_1, m_2, m_3)$, the so-called core tensor. We note that 
the CP decomposition in the previous subsection is a specific Tucker decomposition in which the core
tensor is diagonal, \emph{i.e.}~$G_{ijk} = 0$ unless $i = j = k$, $1 \leq i \leq d_1$, $1 \leq j \leq d_2$, $1 \leq k \leq d_3$.  

Given the above Tucker decomposition, the three modes of $\mathbfcal{T}$ can be expressed as (\emph{cf.}~\cite{Kolda2009}, p.~462). 
\begin{align*}
\mathbfcal{T}_{(1)} = \mathbf{A}^{(1)} \mathbfcal{G}_{(1)} (\mathbf{A}^{(3)} \otimes \mathbf{A}^{(2)}), \qquad 
\mathbfcal{T}_{(2)} = \mathbf{A}^{(2)} \mathbfcal{G}_{(2)} (\mathbf{A}^{(3)} \otimes \mathbf{A}^{(1)}), \qquad 
\mathbfcal{T}_{(3)} = \mathbf{A}^{(3)} \mathbfcal{G}_{(3)} (\mathbf{A}^{(2)} \otimes \mathbf{A}^{(1)}),
\end{align*}
where $\otimes$ denotes the Kronecker product of matrices: for matrices $\M{M}$ and $\M{M}'$ of dimensions
$n \times m$ and $n' \times m'$, respectively, $\M{M} \otimes \M{M}'$ is a matrix of dimensions $(n \cdot n') \times 
(m \cdot m')$ that can be represented compactly as the block matrix $(M_{ij} \M{M}')$ consisting of 
$n \cdot m$ blocks. 

\section{Simulations}
\label{sec:simulation}

The main insight of this paper is that tensor completion provides a simple and flexible approach for combining multiple outcomes that does not rely on strong parametric assumptions. In comparison, traditional methods that rely on a univariate outcome or combine multiple outcomes using strong parametric assumptions may not be sufficiently flexible to study the effect of COVID-19 mandates. To test this insight, we conduct a series of simulation studies.

In our simulations, we denote the primary outcome as $\M{Y} = (Y_{it})$. We only consider one additional outcome $ \M{Z} = (Z_{it})$, and we do not consider covariates and offsets. (More complicated simulations produce similar results.) Let $\mu_{it}$ denote the log-mean of the Negative Binomial distribution (NB) and $\phi^{-1}$ represent the size. We conduct simulations based on the following three distributions, with parameters specified below:

\begin{align*}
& \text{{\bfseries Simulation 1}:} & Y_{it}  &\sim \text{NB} \left ( \text{exp}(\mu_{it} = \theta_i + \eta_t), \; \phi \right )  \\
&  & Z_{it}  &\sim \text{NB} \left ( \text{exp}(\mu_{it} = \theta_i + \eta_t  + \delta), \; \phi \right )\\
& & & \\
& \text{{\bfseries Simulation 2}:} & Y_{it}  &\sim \text{NB} \left ( \text{exp}(\mu_{it} = \theta_i + \eta_t + \zeta_{it}), \; \phi \right )  \\
&  & Z_{it}  &\sim \text{NB} \left ( \text{exp}(\mu_{it} = \theta_i + \eta_t + \zeta_{it} + \delta), \; \phi \right )\\
& & & \\
& \text{{\bfseries Simulation 3}:} & Y_{it}  &\sim \text{NB} \left ( \text{exp}(\mu_{it} = \theta_i + \eta_t + \zeta_{it}), \; \phi \right )  \\
&  & Z_{it}  &\sim \text{NB} \left ( \text{exp}(\mu_{it} = \theta_i + \eta_t + \zeta_{it} + \delta), \; \phi \right )\\
 & & \M{Z}' & = \M{Z}\M{D} 
\end{align*}

In Simulation 1, we draw samples from a log-linear model with row effects $\theta_i$ and column effects $\eta_i$. We assume that $\theta_i \sim \eta_t \sim \text{Normal}(4,1)$, and we set $\delta =-1$ and $\phi=0.01$. In Simulation 2 and 3, we introduce additional complexity to the log-linear model by incorporating interactions $\zeta_{it} \sim \text{Normal}(-(i+t)/30, 0.1)$. In Simulation 3, we also randomly shuffle the columns of the additional outcome to simulate the fact that the additional outcome may lag or lead the outcome of interest. Here, $\M{D}$ is a $T\times T$ shuffle matrix that permutes the order of the columns of the additional outcome matrix $\M{Z}$. We set $N=50$ and $T=8$, and we assume that $10\%$ of the entries in primary outcome are missing at random. To evaluate the model's performance, we use the mean squared error (MSE) where the outcome is measured on the log scale
\begin{equation*}
    \frac{1}{|\mathcal{M}|}\sum_{(i,t)\in\mathcal{M}}\left( \log(\widehat{Y}_{it}) - \log(Y_{it}) \right)^2, 
\end{equation*}
where $\widehat{Y}_{it}$ is the estimator of $Y_{it}$.

We apply the six methods described in Subsection \ref{subsec:results} to the simulated data. The accuracy of all six methods in all three simulation studies is summarized in Table \ref{tab:sim_results}.

\begin{table}[ht]
\centering
\begin{scriptsize}
\begin{tabular}{lc c c}
  \toprule
      Method & \textbf{Simulation 1} & \textbf{Simulation 2} & \textbf{Simulation 3}
        \\[0.5ex]
        \hline
    \specialrule{0em}{2pt}{2pt}
  Log-linear Model (1) & $0.013$ & $0.142$ & $0.142$ \\[0.5ex]
  Log-linear Model (2) & $0.013$ & $0.118$ & $0.135$ \\[0.5ex]
  Log-linear Model (3) & $0.024$ & $0.031$ & $1.980$ \\[0.5ex]
  Matrix Completion (1) & $0.014$ & $0.140$ & $0.140$ \\[0.5ex]
  Matrix Completion (2) & $0.014$ & $0.034$ & $0.140$ \\[0.5ex]
  Tensor Completion  & $0.018$ & $0.050$ & $0.057$ \\[0.5ex]
  \bottomrule
  \specialrule{0em}{3pt}{3pt}
\end{tabular}
\end{scriptsize}
\caption{MSE for Tensor Completion and five comparison methods under three different simulation settings.}
\label{tab:sim_results}
\end{table}

The simulations reveal that all methods perform well when the data follows a log-linear model with row and column effects (Simulation 1). However, when we introduce interactions (Simulation 2), the error of the methods that only use univariate outcomes are more than two times higher than those that use both outcomes. When we shuffle the columns of the additional outcome (Simulation 3), all methods have more than twice the error of the proposed method, Tensor Completion.

We conclude that univariate longitudinal methods, such as the Log-linear Model (1) and Matrix Completion (1), are sensitive to row-column interactions, which represent unobserved confounders that vary by both row and column. It is not sufficient to model the outcome using row effects, column effects, and the additional variable as a covariate. Modeling the additional variable as an outcomes can account for the interaction, but Simulation 3 demonstrates that strong parametric assumptions can produce more error than simply ignoring the additional variable. (\emph{i.e.} Compare Log-linear Model (3) with Log-linear Model (1) and (2) in Simulation 3.) The proposed tensor completion approach allows the researcher to model an additional outcome without making strong parametric assumptions. 
\end{document}